\documentclass[
 aip,
 pop,
 amsmath,
 amssymb,
 reprint,
 floatfix
]{revtex4-2}

\usepackage{amsmath, amssymb, bm}
\usepackage{graphicx}
\usepackage{subcaption}
\usepackage{booktabs}
\usepackage{siunitx}
\usepackage{multirow}
\usepackage{hyperref}
\usepackage{caption}
\usepackage{placeins}
\usepackage{xcolor}

\graphicspath{{figures/}}
\newcommand{\figsinglecol}{0.88\columnwidth}
\newcommand{\figwideone}{0.86\textwidth}
\newcommand{\figtwoperrow}{0.48\textwidth}
\newcommand{\figthreeperrow}{0.31\textwidth}

\hypersetup{
    colorlinks=true,
    linkcolor=blue,
    citecolor=blue,
    urlcolor=blue
}

\AtBeginDocument{%
}

\begin{document}

\title{Likelihood topology and applicability limits of spatial anti-aliasing inversion for folded electron drift instability dispersion in Hall thrusters}
\author{Zilong Peng}
\affiliation{School of Energy Science and Engineering, Harbin Institute of Technology, Harbin 150001, People's Republic of China}
\author{Baisheng Wang}
\affiliation{School of Energy Science and Engineering, Harbin Institute of Technology, Harbin 150001, People's Republic of China}
\author{Zhongping Zhao}
\affiliation{School of Energy Science and Engineering, Harbin Institute of Technology, Harbin 150001, People's Republic of China}
\author{Daren Yu}
\affiliation{School of Energy Science and Engineering, Harbin Institute of Technology, Harbin 150001, People's Republic of China}
\author{Yinjian Zhao}
\email{zhaoyinjian@hit.edu.cn}
\affiliation{School of Energy Science and Engineering, Harbin Institute of Technology, Harbin 150001, People's Republic of China}
\date{June 2, 2026}

\begin{abstract}
The electron drift instability (EDI) is widely recognized as the leading mechanism for anomalous electron transport in Hall thrusters, but its millimeter and sub-millimeter wavelength makes conventional wave-probe analysis strongly limited by spatial aliasing.
A multi-geometry spatial anti-aliasing algorithm provides a route for wave-probe diagnostics to break the spatial Nyquist limit, but its effectiveness for nonmonotonic, folded, multi-resonance EDI dispersion has not been quantified.
Using the numerically solved linear kinetic EDI dispersion relation under typical Hall-thruster discharge parameters as a benchmark, this work evaluates two-dimensional maximum-likelihood spatial anti-aliasing inversion based on two-probe synthetic signals generated for 25 simulated angle--spacing configurations.
For a sampling frequency $f_s=100$~MHz and signal-to-noise ratio (SNR)~=~10, the method breaks the conventional Nyquist wavenumber limit of about 1~krad/m, extends the accessible range to about 22~krad/m, and recovers the dominant likelihood ridges associated with the first- to fifth-order EDI branches.
Parameter scans further show that the statistical realization count $L$ mainly determines the suppression of spurious alias peaks, whereas the frequency resolution $\delta f$ mainly determines local branch separation near folded extrema.
These results provide benchmark sampling and segmentation constraints for multi-geometry wave-probe diagnosis of EDI dispersion under the present synthetic conditions and can guide future experiments.
\end{abstract}


\maketitle

\section{Introduction}

Cross-field electron transport in Hall thrusters is much larger than predicted by classical collision and near-wall transport theories, and this anomalous transport remains a central obstacle to predictive thruster modeling
\cite{boeuf2017,choueiri2001,morozov2000,kaganovich2020}.
Early modeling and experimental studies established the importance of discharge oscillations, anomalous mobility, and thruster-scale transport closures \cite{boeuf1998,boniface2006,hagelaar2002}.
The electron drift instability (EDI) has been identified as the dominant mechanism, receiving sustained attention because it directly couples the $\bm{E}\times\bm{B}$ drift, electron cyclotron resonances, and the ion response
\cite{ducrocq2006,adam2004,adam2008,lafleur2016}.
Nonlinear simulations further show that EDI-related structures and wall-coupled transport can modify saturation and anomalous conductivity \cite{heron2013,janhunen2018nonlinear,janhunen2018ecdi}.
Recent two-dimensional radial--azimuthal PIC simulations further show that radial magnetic-field strength and gradients can shift EDI spectra and anomalous axial mobility \cite{chen2026pst}.
Particle-in-cell (PIC) simulations and kinetic theory can describe EDI growth, saturation, and its connection with anomalous transport, but their validation requires measurements that resolve more than fluctuation amplitude or broadband spectral content.
The wavenumber--frequency dispersion provides a direct test of the Doppler-shifted cyclotron resonance structure and therefore constrains whether the measured fluctuation is consistent with EDI.
An experimental diagnostic capable of resolving this dispersion is consequently a key link between theory, simulation, and real discharge behavior.

EDI diagnostics are difficult because the relevant wavelengths usually lie in the millimeter to sub-millimeter range.
Conventional Beall analysis can infer frequency--wavenumber distributions from cross-spectral phase \cite{beall1982} and has been used in wave diagnostics of Hall thrusters and hollow-cathode plumes \cite{brown2017iepc,brown2018aiaa,jorns2017}.
Coherent light scattering has also observed MHz short-wavelength fluctuations and EDI-like dispersion features in Hall thrusters
\cite{cavalier2013,tsikata2009,lazurenko2008}.
However, probe methods are bounded by the spatial Nyquist limit set by probe spacing, whereas optical methods are limited by accessible wavenumber windows and experimental geometry.
For a typical 3~mm two-probe spacing, the spatial Nyquist wavenumber is $k_\mathrm{Nyq}\approx1$~krad/m, while high-order EDI cyclotron resonances can extend to about 10--22~krad/m.
Thus the high-wavenumber branches that best characterize EDI cyclotron resonance are also the branches most easily hidden by spatial aliasing in conventional probe diagnostics.

The multi-geometry spatial anti-aliasing algorithm proposed by Liu and Jorns provides a new route for this problem
\cite{liu2025,liu2025iepc}.
Related temporal anti-aliasing approaches extend the frequency range of high-speed Hall-thruster imaging diagnostics \cite{liu2026temporal}.
Instead of relying on an apparent wavenumber from one probe spacing, the method uses phase-difference data from multiple probe spacings and angles to build a joint likelihood over candidate two-dimensional wave vectors $(k_y,k_z)$, where $k_y$ and $k_z$ denote the axial and azimuthal wave-vector components, respectively.
The true wave vector can then be selected from multiple aliased candidates.
Liu and Jorns previously verified this anti-aliasing strategy numerically and experimentally in hollow-cathode ion-acoustic-wave (IAW) turbulence, where the dispersion is nearly linear and monotonic \cite{liu2025,liu2025iepc}.
Here the IAW result is used only as a prior diagnostic demonstration, because EDI has distinct wave physics and a nonlinear, folded dispersion near multiple cyclotron resonances.
Near the resonance condition $k_zV_d\approx n\omega_{ce}$, where $V_d$ is the electron drift velocity, $n$ is the resonance order, and $\omega_{ce}$ is the electron cyclotron angular frequency, the EDI branch becomes nonmonotonic and multi-valued in frequency--wavenumber space.
Several nearby or separated $k_z$ candidates can therefore contribute to the same frequency bin, which creates a more demanding test for likelihood-based anti-aliasing.
Whether the spatial anti-aliasing algorithm remains stable in this EDI regime, how many statistical realizations are needed, and how frequency resolution affects folded-region accuracy are therefore questions that must be answered before experimental wave-probe diagnostics are designed for EDI.

This work uses synthetic signals generated from folded EDI dispersion as a benchmark and evaluates the feasibility of wave-probe diagnostics within the multi-geometry anti-aliasing framework.
Specifically, the theoretical EDI dispersion is solved using the Hall-thruster parameters of Cavalier et al. \cite{cavalier2013}, two-probe synthetic signals are generated from this dispersion as ground truth, and a two-dimensional maximum-likelihood inversion is performed using 25 angle--spacing configurations.
The benchmark prescribes azimuthally propagating modes with $k_y=0$, while possible axial components and oblique EDI propagation are treated as experimental extensions.
By comparing two schemes, increasing the realization count $L$ at fixed frequency resolution and reducing the frequency resolution $\delta f$ at fixed $L$, this paper separates the effects of statistical convergence, frequency-bin mixing, and multi-peak competition.
The remainder of this paper is organized as follows.
Section~II introduces Beall analysis and the spatial anti-aliasing method.
Section~III presents the EDI benchmark, synthetic signals, and test cases.
Section~IV reports the inversion results and failure mechanisms.
Section~V discusses implications for real EDI wave-probe experiment design.
Finally, Sec.~VI summarizes the main conclusions.

\section{Theoretical background and analysis method}

\subsection{EDI dispersion relation}

In a Hall thruster, the background axial electric field and radial magnetic field produce an azimuthal electron $\bm{E}\times\bm{B}$ drift.
The corresponding drift speed is written as $V_d=E_0/B_0$, where $E_0$ and $B_0$ are the background electric-field and magnetic-field strengths, respectively.
The coordinate system used here follows the Liu--Jorns convention for the local Hall-thruster geometry: $x$ is radial and approximately parallel to the applied magnetic field, $y$ is axial, and $z$ is azimuthal.
For electrostatic perturbations in this crossed-field plasma, the EDI dispersion relation is determined by the zeros of the dielectric function

\begingroup
\small
\begin{equation}
  \begin{aligned}
  \varepsilon(\omega, \bm{k})
  =&\ 1 + k^2\lambda_{De}^2
  + g\!\left(
    \frac{\omega - k_z V_d}{\omega_{ce}},
    k_\perp^2\rho_e^2,
    k_x^2\rho_e^2
  \right) \\
  &- \frac{k^2\lambda_{De}^2\,\omega_{pi}^2}{\omega^2}
  = 0 .
  \end{aligned}
  \label{eq:dispersion}
\end{equation}
\endgroup

Equation~\eqref{eq:dispersion} uses a Poisson-form normalization equivalent to that used by Ducrocq, Cavalier, and co-workers \cite{ducrocq2006,cavalier2013}.
The terms in Eq.~\eqref{eq:dispersion} represent Debye shielding, the magnetized electron cyclotron response, and the cold-ion response, respectively.
Here $\varepsilon$ is the dielectric function, $\omega$ is the complex angular frequency, $\bm{k}$ is the wave vector, $k=|\bm{k}|$ is the wavenumber magnitude, and $k_x$, $k_y$, and $k_z$ are the radial, axial, and azimuthal wavenumber components.
The quantities $\lambda_{De}$, $\rho_e$, $\omega_{ce}$, and $\omega_{pi}$ denote the electron Debye length, electron Larmor radius, electron cyclotron angular frequency, and ion plasma angular frequency, respectively.
Because the radial direction is taken to be approximately along the magnetic field, $k_x$ is the parallel component and $k_\perp$ denotes the component perpendicular to the magnetic field.
The function $g$ is the Gordeev function, which encodes electron cyclotron motion through a series expansion in the Doppler-shifted normalized frequency $\Omega=(\omega-k_zV_d)/\omega_{ce}$.
In the numerical solution below, $g$ is evaluated with the same series representation used in the EDI calculations of Ducrocq and Cavalier, with the harmonic sum truncated after convergence of the dielectric residual.
The last term gives the cold-ion susceptibility.
When an axial ion beam is retained, this response is generally written with $(\omega-k_y v_p)^{-2}$, where $v_p$ is the axial ion-beam velocity; in the present benchmark, $k_yv_p=0$, so the response reduces to $\omega^{-2}$.
Similar EDI dielectric forms and analytic or numerical approximations under Hall-thruster parameters have also been used in later theory and PIC comparison studies \cite{lafleur2016,boeuf2018}.
Solving Eq.~\eqref{eq:dispersion} in the complex $\omega$ plane gives the real frequency $\omega_R$ and growth rate $\gamma$.
The present work uses the resulting dependence on the azimuthal wavenumber $k_z$ as the ground-truth dispersion for the spatial anti-aliasing test.

When the cyclotron resonance condition $k_zV_d\approx n\omega_{ce}$ is satisfied, the dispersion curve develops local extrema and folds, and the growth rate can exhibit localized peaks.
This behavior produces the multi-resonance structure of EDI.
The structure corresponds to the coupling of electron Bernstein/cyclotron harmonics, Doppler shifted by the $\bm{E}\times\bm{B}$ drift, to the low-frequency ion response
\cite{ducrocq2006,cavalier2013,boeuf2018}.
The specific dispersion shape used in this work is presented in \hyperref[sec:plasma_parameters]{Sec.~III.A}.

\subsection{Conventional Beall analysis and its aliasing limit}

The two-probe cross-spectral method proposed by Beall et al. \cite{beall1982} is a classical technique for estimating local frequency--wavenumber spectra in plasma-wave diagnostics.
It has later been widely used for dispersion measurements in Hall-thruster channels and near-field plumes \cite{brown2017iepc,brown2018aiaa} and in hollow-cathode plumes \cite{jorns2017,liu2024iepc}.
In an experiment, two probes separated by $\Delta r$ commonly measure time-varying ion-saturation currents; in the present synthetic benchmark, the same procedure is represented by two spatially separated time signals.
A propagating plasma wave introduces a phase difference between the two signals.
The apparent wavenumber projected along the probe-separation direction can then be estimated from the phase of the cross spectrum:

\begingroup
\small
\begin{equation}
  k_\mathrm{Beall}(f) = \frac{1}{\Delta r}
  \angle\!\left(F_2(f) \cdot F_1^*(f)\right)
  \label{eq:beall}
\end{equation}
\endgroup

Here $k_\mathrm{Beall}$ is the baseline-projected apparent wavenumber inferred by the Beall method, $f$ is frequency,
$F_1(f)$ and $F_2(f)$ are the Fourier coefficients of the two probe signals at frequency $f$,
the superscript $*$ denotes complex conjugation, and $\angle(\cdot)$ denotes the complex argument.
With the cross-spectrum convention $F_2F_1^*$ used here, the sign of $k_\mathrm{Beall}$ follows the ordering of the two probes and the definition of the baseline vector; reversing the probe order reverses the plotted signed wavenumber but does not change the Nyquist limit or aliasing analysis.
Building a two-dimensional frequency--wavenumber histogram over multiple realizations gives a Beall dispersion map.
This map is a power-weighted distribution of cross-spectral phase estimates rather than a unique deterministic dispersion curve.
Because the arctangent phase is restricted to $(-\pi,\pi]$, however, the inferred wavenumber is strictly bounded by the spatial Nyquist limit,

\begin{equation}
  |k_\mathrm{Beall}| \leq k_\mathrm{Nyq} = \frac{\pi}{\Delta r}.
  \label{eq:nyquist}
\end{equation}

When the wavenumber projected onto the probe baseline exceeds $k_\mathrm{Nyq}$, the phase wraps.
The true wavenumber is then folded back into the Nyquist interval
$(-k_\mathrm{Nyq},k_\mathrm{Nyq}]$, producing spatial aliasing.
For the azimuthal baselines emphasized in this EDI benchmark, this projected wavenumber is dominated by $k_z$.
For the 3--5~mm probe spacings used here, $k_\mathrm{Nyq}$ is about 0.63--1.05~krad/m, whereas the target EDI wavenumber range reaches 22~krad/m.
Thus Beall analysis is used here as an aliasing baseline: it shows what a conventional two-probe diagnostic would observe and identifies the spatial-aliasing barrier for high-order EDI resonances.
Jorns et al. \cite{brown2019} attempted to manually de-alias Beall maps using their periodic structure; that procedure is most straightforward when a single dominant mode with nearly linear dispersion is present, whereas the present EDI benchmark contains nonlinear folded dispersion with multiple resonance orders.
Similar phase-wrapping and spatial-aliasing issues directly motivated the spatial anti-aliasing method \cite{liu2025,liu2025iepc}.

\subsection{Spatial anti-aliasing algorithm}

\begin{figure}[t]
  \centering
  \includegraphics[width=\figsinglecol]{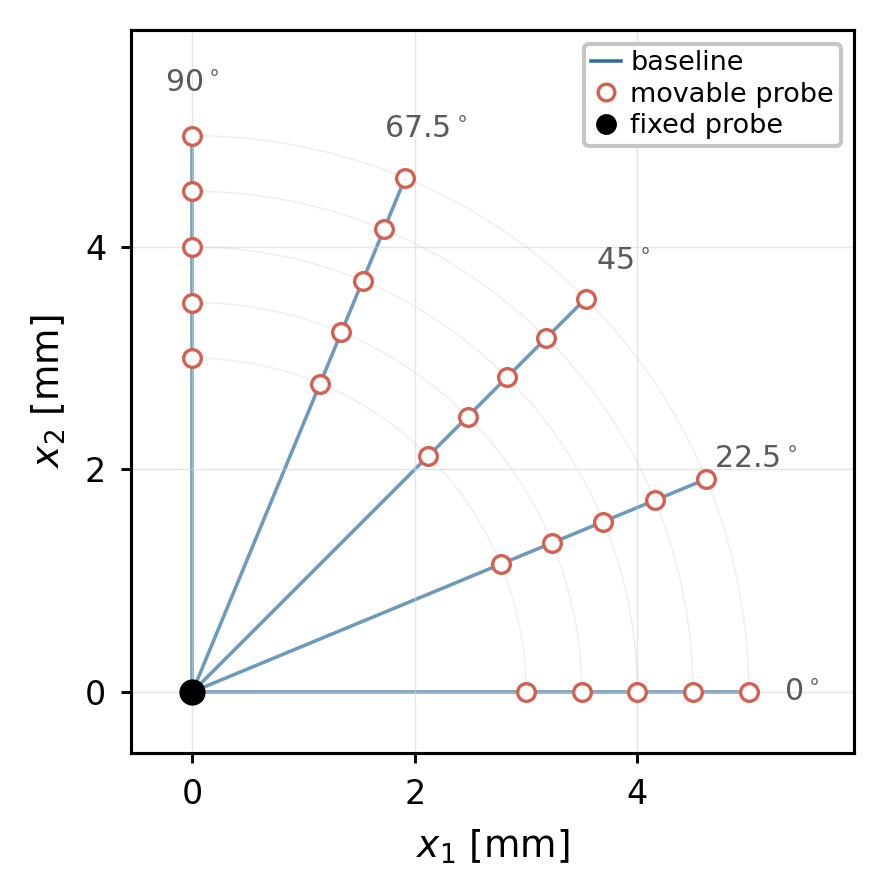}
  \caption{Schematic of the multi-geometry probe configuration.
           The fixed probe is at the origin, while the movable probe forms several baselines with different spacings and angles.
           Different baselines impose different phase-projection constraints and therefore provide complementary information for joint-likelihood inversion.}
  \label{fig:liu_probe_geometries}
\end{figure}

The core idea of the spatial anti-aliasing algorithm of Liu and Jorns \cite{liu2025} is that changing probe spacing and orientation changes the aliased phase constraint in a geometry-specific way.
By jointly analyzing phase differences from many probe geometries through Bayesian inference, the true wave vector can be recovered from aliased data.
Figure~\ref{fig:liu_probe_geometries} shows a typical multi-geometry probe array.
The fixed probe is located at the origin, and the movable probe forms baselines with different lengths and angles.
Each baseline constrains the projection of the wave vector along that direction.
Baselines with different directions and lengths therefore provide complementary constraints, which form the geometric basis for breaking the Nyquist limit of a single probe pair.
This multi-geometry strategy differs from manual de-aliasing based on periodic Beall-map structures because it tests whether a candidate wave vector remains phase-consistent across all configurations
\cite{brown2017iepc,liu2025,liu2025iepc,brown2018aiaa}.

\begin{figure*}[t]
  \centering
  \includegraphics[width=\figwideone]{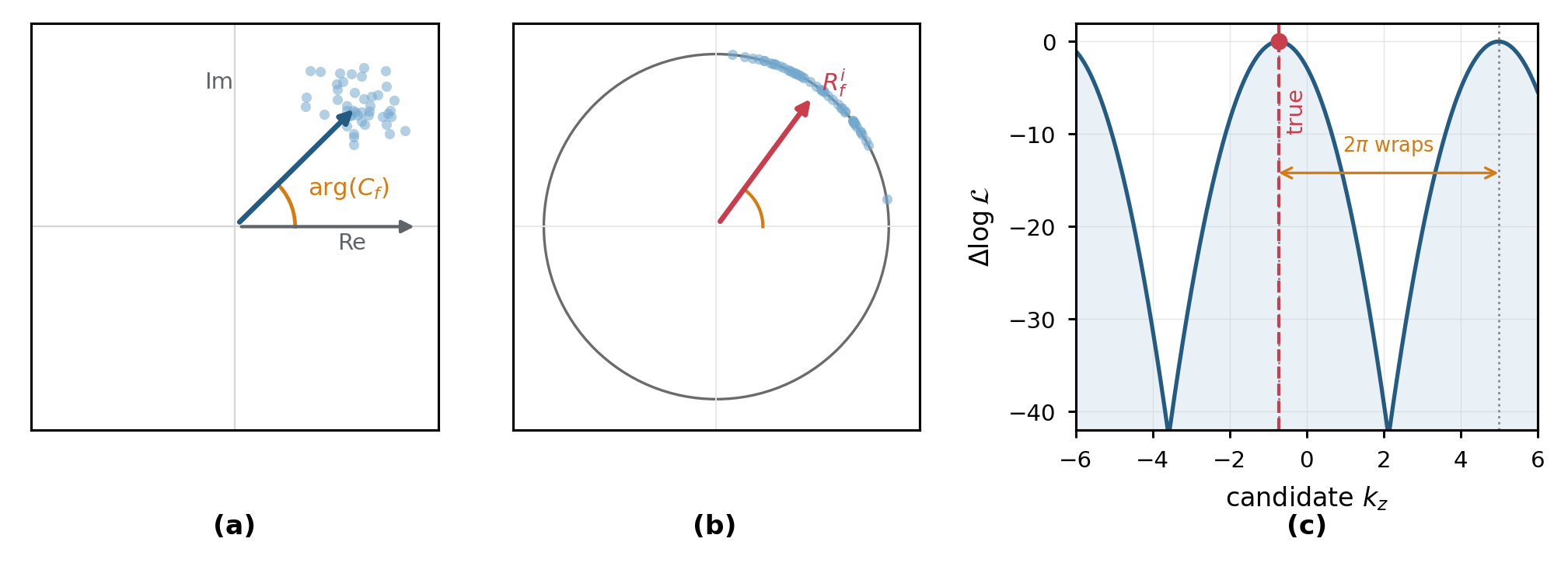}
  \caption{Schematic of phase statistics and single-geometry likelihood construction.
           (a) The argument of the cross spectrum $C_f=S_{2,f}S_{1,f}^*$ gives a single phase-difference sample.
           (b) Phase samples from multiple realizations are mapped onto the unit circle; the direction of the mean phasor gives the circular mean and its length reflects phase concentration.
           (c) For one probe geometry, phase wrapping makes several candidate $k_z$ values produce identical or similar phase residuals, yielding periodic likelihood peaks.}
  \label{fig:phase_statistics_schematic}
\end{figure*}

\textbf{Phase statistics.}
For $n_\mathrm{geo}$ probe geometries, each two-probe signal is first divided into $L$ realizations, equivalently $L$ non-overlapping time windows.
Here $n_\mathrm{geo}$ is the number of probe geometries, and $L$ is the statistical realization count.
For geometry $i$ and realization $l$, the Fourier coefficients of the two probe signals are denoted by $S_{1,f}^{i,(l)}$ and $S_{2,f}^{i,(l)}$.
Here $i$ is the geometry index, $l$ is the realization index, and the subscript $f$ denotes the frequency bin.
The cross spectrum is
\begin{equation}
  C_f^{i,(l)} =
  S_{2,f}^{i,(l)}
  \left(S_{1,f}^{i,(l)}\right)^* ,
  \label{eq:cross_spectrum}
\end{equation}
and the phase-difference sample is
\begin{equation}
  \delta\Theta_f^{i,(l)} =
  \arg C_f^{i,(l)} .
  \label{eq:phase_sample}
\end{equation}
Because $\delta\Theta$ is restricted to $(-\pi,\pi]$, ordinary linear averaging produces errors near $\pm\pi$.
We therefore first compute the mean phasor on the unit circle,
\begin{equation}
  Z_f^i =
  \frac{1}{L}\sum_{l=1}^{L}
  \exp\!\left(i\delta\Theta_f^{i,(l)}\right),
  \label{eq:mean_resultant}
\end{equation}
and then use its argument as the circular mean,
\begin{equation}
  \overline{\Delta\Theta}_f^i =
  \arg Z_f^i .
  \label{eq:circular_mean_method}
\end{equation}
The phase variance is computed after wrapping each sample around the circular mean:
\begin{equation}
  \begin{aligned}
  (\sigma^2)_f^i =
  \frac{1}{L-1}\sum_{l=1}^{L}
  &\left[
    \operatorname{wrap}_\pi(\Delta\theta_f^{i,(l)})
  \right]^2
  + \epsilon,\\
  \Delta\theta_f^{i,(l)}
  =&\ \delta\Theta_f^{i,(l)}
  - \overline{\Delta\Theta}_f^i .
  \end{aligned}
  \label{eq:phase_variance_method}
\end{equation}
where $\operatorname{wrap}_\pi(\cdot)$ maps an angle back to $(-\pi,\pi]$, and $\epsilon=10^{-12}$ is a numerical regularization.
Figure~\ref{fig:phase_statistics_schematic} illustrates the relation among cross-spectral phase extraction, circular averaging, and periodic single-geometry likelihood peaks.

\textbf{Likelihood function.}
This work uses the two-dimensional wave-vector search form of the spatial anti-aliasing method.
The candidate wave vector in the probe measurement plane is $\bm{k}=(k_y,k_z)$.
The baseline of geometry $i$ is $\Delta\bm{r}^i=(\Delta y^i,\Delta z^i)$, and its predicted phase difference is
$-\bm{k}\cdot\Delta\bm{r}^i=-(k_y\Delta y^i+k_z\Delta z^i)$.
Because the observed and predicted phases are both equivalent modulo $2\pi$, their comparison uses a wrapped phase residual:
\begin{equation}
  \begin{aligned}
  r_f^i(k_y,k_z) =
  \operatorname{wrap}_\pi
  \left[
  \overline{\Delta\Theta}_f^i
  + k_y\Delta y^i
  + k_z\Delta z^i
  \right].
  \end{aligned}
  \label{eq:wrapped_residual}
\end{equation}
Assuming independent geometries and approximating the wrapped phase residual by a Gaussian penalty, the single-geometry log likelihood is
\begin{equation}
  \ln\mathcal{L}_{f}^{i}(k_y,k_z)
  =
  -\frac{1}{2}
  \frac{\left[r_f^i(k_y,k_z)\right]^2}{(\sigma^2)_f^i}.
  \label{eq:single_loglik}
\end{equation}
The joint log likelihood over all geometries is
\begin{equation}
  \ln\mathcal{L}_{f}(k_y,k_z) = \sum_{i=1}^{n_\mathrm{geo}}
  \ln\mathcal{L}_{f}^{i}(k_y,k_z).
  \label{eq:joint_loglik}
\end{equation}

\begin{figure*}[t]
  \centering
  \includegraphics[width=\figwideone]{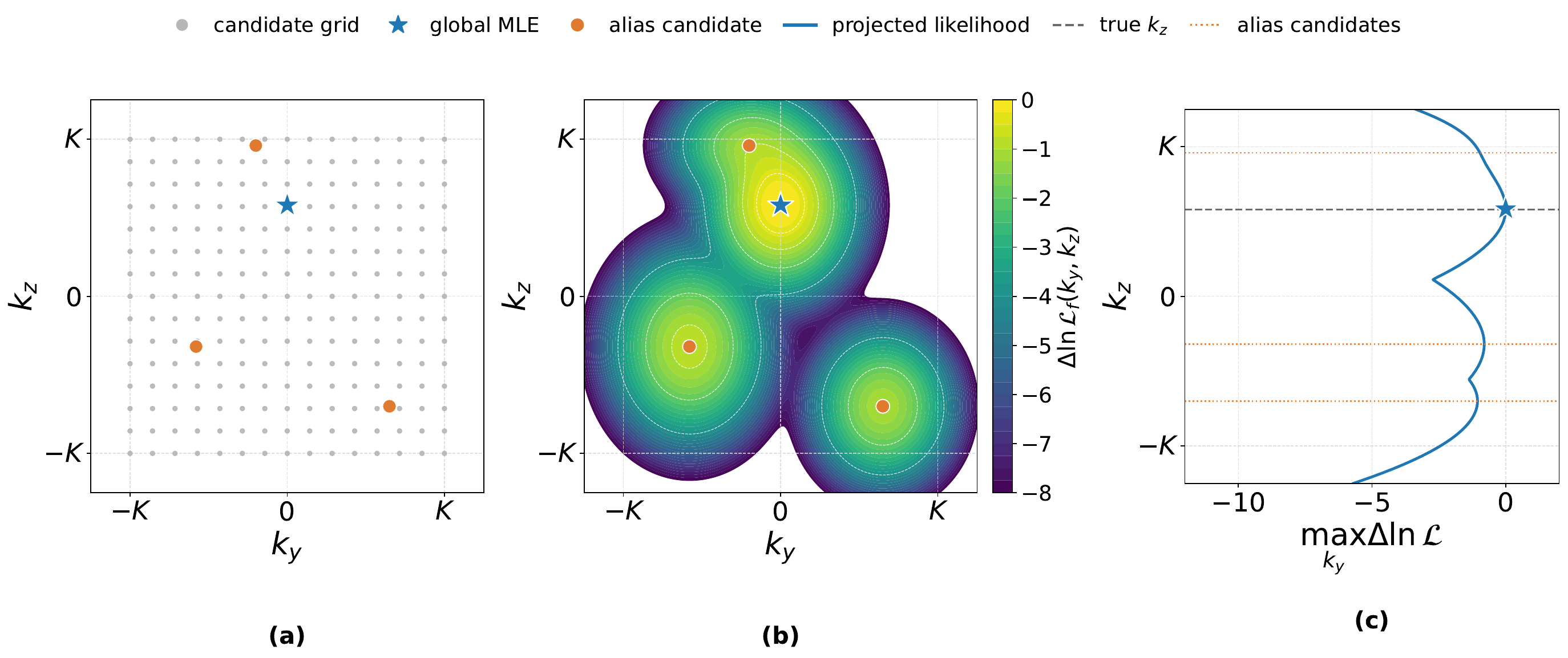}
  \caption{Schematic of two-dimensional grid-search MLE and $k_z$ projection.
           (a) For each frequency $f$, the algorithm enumerates candidate wave vectors on the two-dimensional grid $(k_y,k_z)$ and evaluates the joint likelihood.
           The blue star marks the global maximum-likelihood point $(\hat{k}_y,\hat{k}_z)$, and orange points denote possible alias candidates.
           (b) Multi-geometry joint-likelihood topology at one frequency, where true and alias peaks compete in the two-dimensional wave-vector plane.
           (c) The projected profile $\ln\mathcal{L}^{\rm proj}_f(k_z)=\max_{k_y}\ln\mathcal{L}_f(k_y,k_z)$ is obtained by maximizing over $k_y$.}
  \label{fig:mle_2d_projection_schematic}
\end{figure*}

\begin{figure*}[t]
  \centering
  \includegraphics[width=\figwideone]{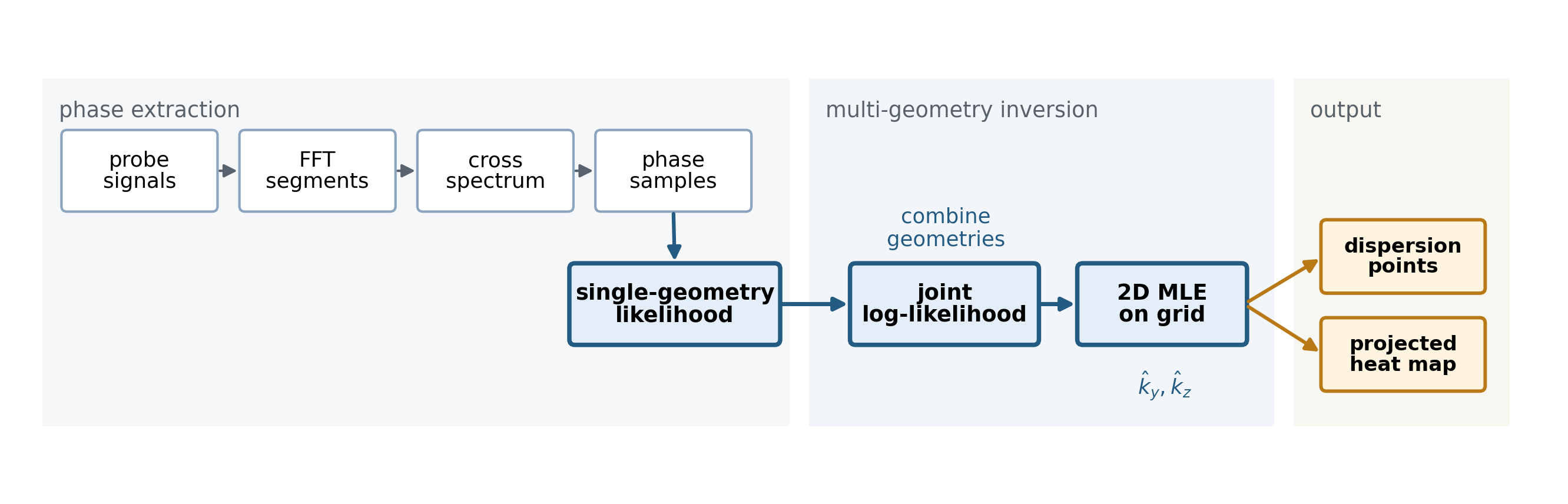}
  \caption{Input-output workflow of the spatial anti-aliasing algorithm.
           Cross-spectral phase samples are converted into single-geometry likelihoods through circular statistics.
           After multi-geometry combination, two-dimensional MLE gives the wave-vector estimate and further produces the $f$--$k_z$ dispersion points and projected likelihood heat maps.}
  \label{fig:anti_aliasing_workflow}
\end{figure*}

\textbf{Maximum likelihood estimation (MLE).}
For each frequency $f$, Eq.~\eqref{eq:joint_loglik} is evaluated over a prescribed two-dimensional candidate grid
$\mathcal{G}_{k_y}\times\mathcal{G}_{k_z}$, and the estimated wave vector is
\begin{equation}
  (\hat{k}_y,\hat{k}_z)(f) =
  \arg\max_{(k_y,k_z)\in\mathcal{G}_{k_y}\times\mathcal{G}_{k_z}}
  \ln\mathcal{L}_{f}(k_y,k_z).
  \label{eq:mle_kxky}
\end{equation}
Thus the grid search is the numerical implementation of MLE: candidate $(k_y,k_z)$ values are enumerated, their joint likelihoods are computed, and the maximum-likelihood grid point is selected as $(\hat{k}_y,\hat{k}_z)(f)$.
The synthetic EDI benchmark used below is azimuthally propagating, so the prescribed ground truth has $k_y=0$ and the diagnostic target is the azimuthal wavenumber $k_z$.
The figures therefore mainly show the projection of the two-dimensional MLE onto the $k_z$ direction.
The direct output of the two-dimensional search is $(\hat{k}_y,\hat{k}_z)$, whereas the $f$--$k_z$ dispersion scatter plots display the $\hat{k}_z$ component.
For likelihood heat maps, this work uses the same two-dimensional search and then projects the result onto $k_z$.
For this projection, the two-dimensional joint likelihood is maximized along the $k_y$ direction to define the projected likelihood profile
\begin{equation}
  \ln\mathcal{L}^{\rm proj}_f(k_z)
  =
  \max_{k_y}\left[\ln\mathcal{L}_{f}(k_y,k_z)\right].
  \label{eq:projected_likelihood}
\end{equation}
This profile is used for constructing $f$--$k_z$ likelihood heat maps and analyzing alias ghosts.
Figure~\ref{fig:mle_2d_projection_schematic} illustrates the relation between the two-dimensional grid search and the $k_z$ projection.
The scatter points and heat maps in the $f$--$k_z$ plots are therefore $k_z$-projected representations of the two-dimensional $(k_y,k_z)$ joint likelihood after maximizing over $k_y$.
The workflow in Fig.~\ref{fig:anti_aliasing_workflow} summarizes the input-output relation among these quantities.

\begin{figure*}[t]
  \centering
  \includegraphics[width=\figwideone]{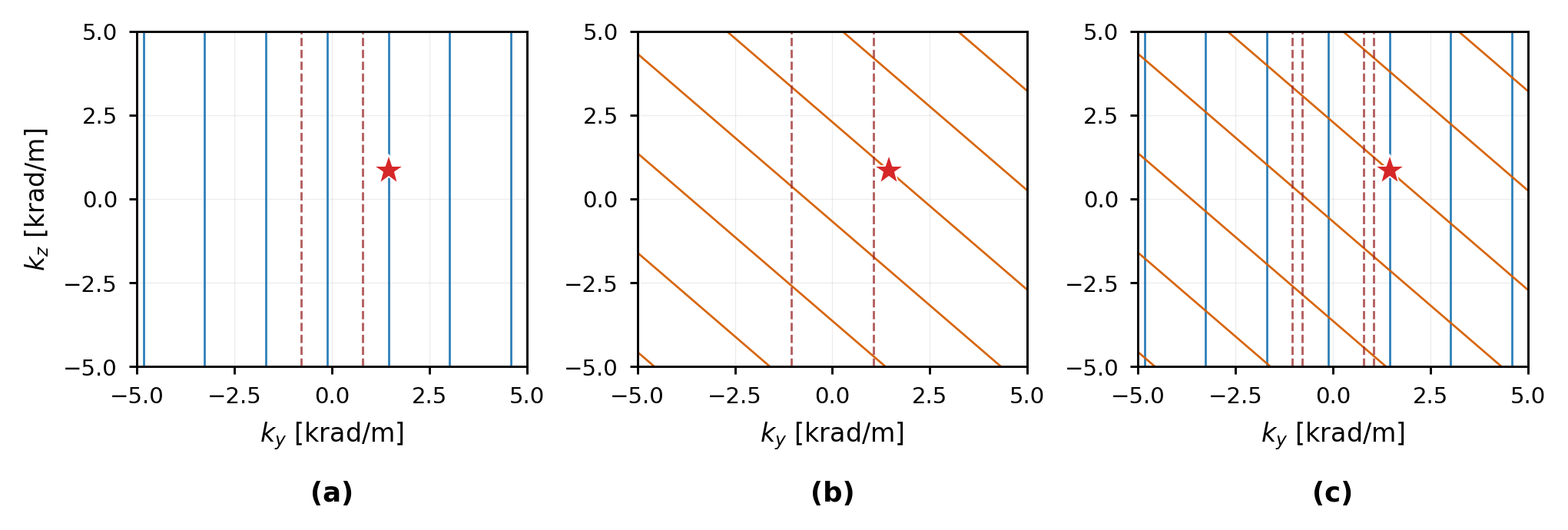}
  \caption{Geometric constraint mechanism of the spatial anti-aliasing algorithm.
           Blue and orange ridges show the log-likelihood constraints from the two probe geometries; red star: true wave vector $\vec{k}_{\rm true}$.
           (a) Single-geometry likelihood for $\Delta r=4.0$~mm, $\alpha=0^\circ$.
           (b) Single-geometry likelihood for $\Delta r=3.0$~mm, $\alpha=45^\circ$.
           (c) Superposition of the two constraints.
           White dashed lines: single-baseline aliasing boundary $|k|=k_N=\pi/\Delta r$.}
  \label{fig:liu_two_config_likelihood}
\end{figure*}

\textbf{Anti-aliasing mechanism.}
For a single geometry, phase wrapping creates a family of candidate wave vectors satisfying
$-\bm{k}\cdot\Delta\bm{x}^i=\mathrm{const.}\pmod{2\pi}$.
These candidates have identical or similar likelihood values and form indistinguishable spurious maxima.
After multiple geometries with different spacings and angles are included, the true wave vector remains consistent across configurations, while spurious maxima are suppressed in the joint likelihood.
This consistency requirement allows the method to exceed the single-baseline Nyquist limit in principle.
Figure~\ref{fig:liu_two_config_likelihood} illustrates this mechanism for two-dimensional wave-vector inversion.
A single geometry gives periodic high-likelihood stripes in which the true wave vector and aliased candidates are degenerate.
Changing the probe spacing or angle changes the stripe position and orientation, so different geometries compress the continuous stripes into discrete candidate intersections.
With two geometries, several aliased intersections may remain.
As more geometries are added, spurious intersections usually fail to satisfy all phase constraints, whereas the true wave vector remains consistent in the joint likelihood.
The numerical experiments below use the same two-dimensional $(k_y,k_z)$ grid search.
If the two-dimensional MLE identifies the correct azimuthal EDI branch, its peak should concentrate near $k_y\approx0$, and the $f$--$k_z$ projection can be compared directly with the theoretical EDI dispersion.

\section{Synthetic benchmark and inversion setup}

This section establishes a controlled benchmark for evaluating the diagnostic algorithm.
Unlike a real experiment, the numerical signal has a known ground-truth dispersion relation, so the spatial anti-aliasing algorithm can be judged directly by whether it recovers the correct branches.
The simulation consists of three steps:
first solving the EDI dispersion relation for a prescribed Hall-thruster parameter set, then synthesizing two-probe signals from this dispersion, and finally applying multi-geometry MLE inversion under different sampling and segmentation conditions.

\subsection{Plasma parameters}
\label{sec:plasma_parameters}

The simulation uses the exit-plane parameters of the Snecma 5~kW PPSX000-VR Hall thruster reported by Cavalier et al.~\cite{cavalier2013} from collective light-scattering measurements (Table~\ref{tab:plasma}).
These parameters have been repeatedly adopted in later EDI experiments and theoretical studies \cite{brown2019,brown2024phd,tsikata2009,brown2018aiaa,brown2021aiaa}, ensuring that the synthetic dispersion is physically representative and directly comparable with prior work.

\begin{table}[!tb]
  \centering
  \scriptsize
  \setlength{\tabcolsep}{2pt}
  \caption{Plasma parameters from Cavalier 2013 at the thruster exit plane.}
  \label{tab:plasma}
  \begin{tabular}{llll}
    \toprule
    Parameter & Symbol & Value & Unit \\
    \midrule
    Plasma density             & $n_0$             & $2.0\times10^{17}$ & m$^{-3}$ \\
    Electron temperature       & $T_e$             & 25                 & eV       \\
    Magnetic field             & $B_0$             & 15                 & mT       \\
    Axial electric field       & $E_0$             & $10^{4}$           & V/m      \\
    $E\times B$ drift speed    & $V_d$             & $\approx667$       & km/s     \\
    Electron cyclotron frequency & $f_{ce}$        & $\approx420$       & MHz      \\
    Ion plasma frequency       & $f_{pi}$          & $\approx52$        & MHz      \\
    Debye length               & $\lambda_{De}$    & $\approx83$        & $\mu$m   \\
    Electron Larmor radius     & $\rho_e$          & $\approx0.80$      & mm       \\
    Normalized radial wavenumber & $k_x\lambda_{De}$ & 0.01             & --       \\
    \bottomrule
  \end{tabular}
\end{table}

\begin{figure}[t]
  \centering
  \includegraphics[width=\figsinglecol]{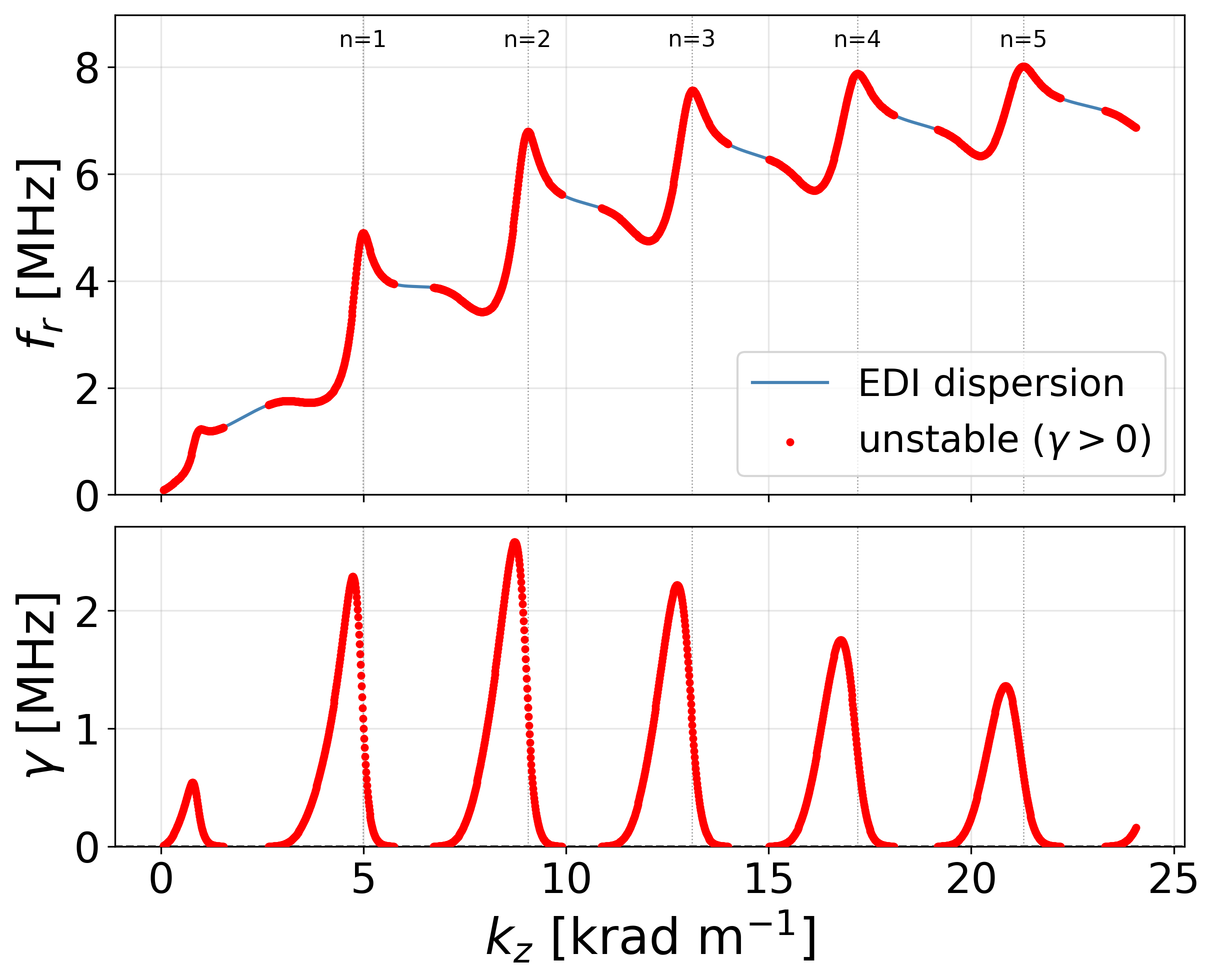}
  \caption{EDI dispersion relation for the Cavalier 2013 parameters: real frequency $f_r$ (top) and growth rate $\gamma$ (bottom) versus azimuthal wavenumber $k_z$.
           Red points: unstable modes ($\gamma>0$); blue line: stable intervals ($\gamma\leq0$).
           The dispersion folds at five cyclotron resonances, where the folded extrema satisfy $v_g=0$.}
  \label{fig:dispersion}
\end{figure}

Although $f_{ce}\approx 420$~MHz, the EDI real frequencies appear in the MHz range, as set by the Doppler-shifted ion response in Eq.~\eqref{eq:dispersion} and shown in Fig.~\ref{fig:dispersion}.
This case is chosen to provide a stringent test:
the dispersion simultaneously contains multiple cyclotron resonances, high-wavenumber extension, and folded multi-valued regions.
The radial wavenumber is set to $k_x\lambda_{De}=0.01$ to preserve clear cyclotron resonances and folded structures, focusing the test on the ability of spatial anti-aliasing to handle high-wavenumber, multi-valued dispersion.
Under this condition, probe phase differences undergo many $2\pi$ wraps, and one frequency bin near folded extrema can contain several nearby $k_z$ branches, simultaneously testing statistical convergence, frequency resolution, and peak selection.

Figure~\ref{fig:dispersion} shows the EDI dispersion relation obtained for these parameters.
The first five cyclotron resonances lie within 0--22~krad/m, far beyond the Nyquist limit of conventional Beall analysis.
The two folded extrema near the first resonance are broad and flat, corresponding to a wide region with $v_g=\partial\omega_R/\partial k_z\approx0$, whereas higher-order resonances have sharper local extrema.
This difference provides the physical basis for the convergence thresholds and local folded-point errors discussed below.

Because the EDI dispersion reduces to an ion-acoustic form in the limit where the magnetized electron cyclotron response in Eq.~\eqref{eq:dispersion} is weak, the IAW frequency provides a reliable initial guess across the full $k_z$ range.
For each prescribed $k_z$, the IAW dispersion relation
$\omega_0=k_zc_s/\sqrt{1+k^2\lambda_{De}^2}$
is used as the initial guess, and Muller's method is used to solve Eq.~\eqref{eq:dispersion} in the complex $\omega$ plane.
The convergence criterion is $|\varepsilon(\omega,\bm{k})|<10^{-8}$.
Here $\omega_0$ is the initial angular-frequency guess, and $c_s$ is the ion sound speed.

The $k_z$ scan covers 0 to 22~krad/m with about 2000 sampling points and is refined near resonant peaks to resolve folded structures.
All converged solutions are retained, including stable intervals with $\gamma\leq0$.
Keeping the stable intervals makes the benchmark dispersion continuous through folded valleys and allows the algorithm to be tested on the folded geometry itself.
These stable-interval points are therefore prescribed spectral components used to test inversion continuity through folded regions; their inclusion should not be interpreted as a claim that linearly stable modes would grow spontaneously in an experiment.

\FloatBarrier
\subsection{Synthetic signal generation}

\subsubsection{Signal model}

The synthetic signal is a superposition of all converged modes plus white noise.
For probe geometry $i$ with azimuthal projected spacing $\Delta z^i$, the two probe signals are

\begingroup
\footnotesize
\begin{align}
  I_1(t) &= \sum_{n=1}^{N_\mathrm{mode}} A_n
             \sin(2\pi f_n t + \phi_n^{(l)}) + \xi_1(t)
             \label{eq:sig1} \\
  I_2(t) &= \sum_{n=1}^{N_\mathrm{mode}} A_n
             \sin\!\left(2\pi f_n t + \phi_n^{(l)}
             - k_{z,n}\Delta z^i\right)
             + \xi_2(t)
             \label{eq:sig2}
\end{align}
\endgroup

Here $t$ is time, $I_1(t)$ and $I_2(t)$ are the two probe time-domain signals, $N_\mathrm{mode}$ is the total number of modes, $A_n$ is the amplitude of mode $n$,
$f_n$ is the mode frequency, and $k_{z,n}$ is the mode azimuthal wavenumber.
$\phi_n^{(l)}$ is the initial phase of mode $n$ in realization $l$ and is independently sampled from a uniform distribution over $[0,2\pi)$ for each realization.
The terms $\xi_1(t)$ and $\xi_2(t)$ are independent white Gaussian noises.
Because the prescribed EDI modes propagate purely in the azimuthal direction ($k_{y,n}=0$), only the azimuthal projection $\Delta z^i$ of the baseline enters the phase difference, even when a geometry has $\Delta y^i\neq 0$.
The ground truth for the two-dimensional inversion is therefore $k_y=0$.

The signal model retains the elements required to test spatial anti-aliasing inversion: a known EDI dispersion, finite sampling, white noise, and spatial phase wrapping.
Other features of real Hall-thruster fluctuations and measurements --- finite probe size, sheath-scale collection, probe perturbation, discharge nonstationarity, nonlinear saturation, mode broadening, and cross-scale energy transfer \cite{brown2023prl,brown2023pre,brown2019,brown2018aiaa,brown2021aiaa} --- fall outside the scope of the present numerical benchmark and are addressed in experimental validation.

\begin{figure}[t]
  \centering
  \includegraphics[width=\figsinglecol]{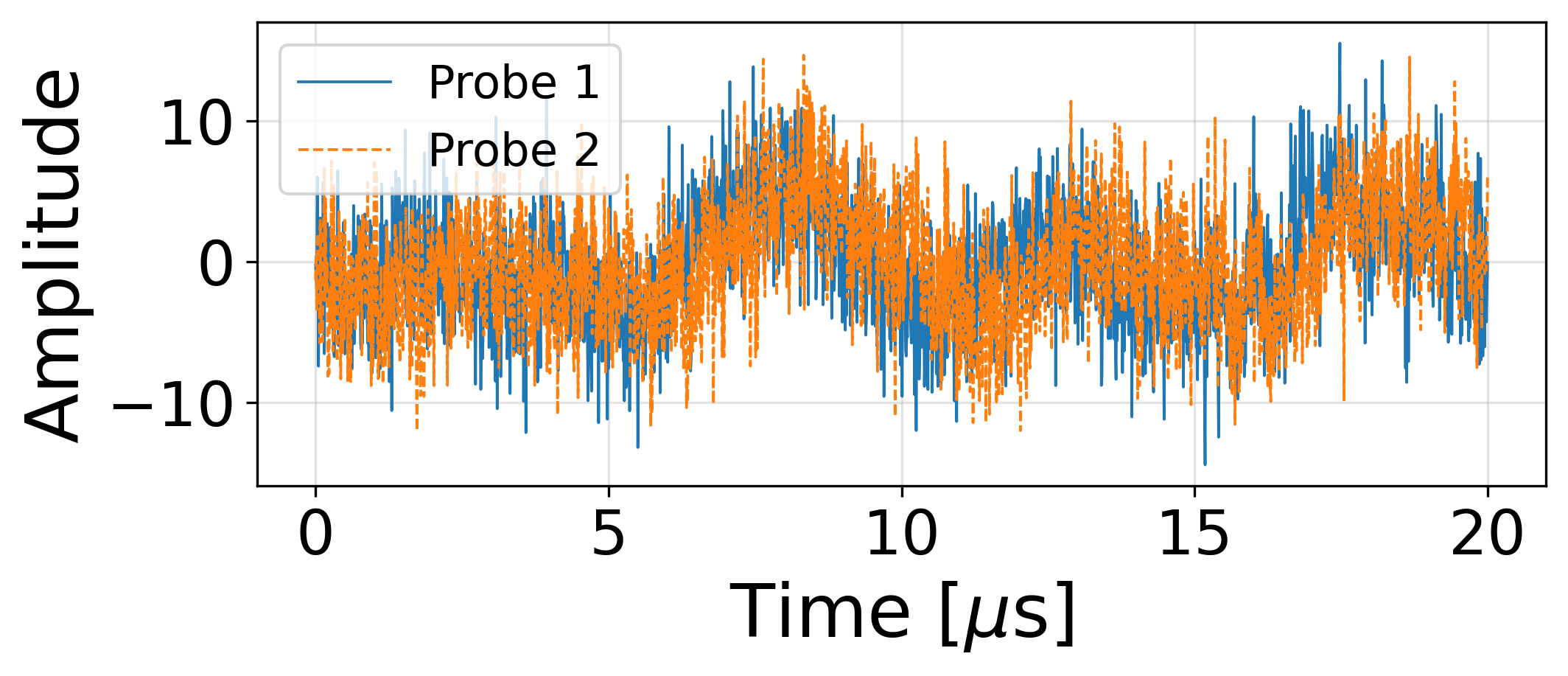}
  \caption{Example raw two-probe waveform for the $|r_p|=3.0$~mm, $\alpha=90^\circ$ geometry over the first 20~$\mu$s.}
  \label{fig:probe_signal}
\end{figure}

Figure~\ref{fig:probe_signal} shows a representative segment of the raw two-probe waveform.
The two signals have comparable amplitudes, and their phase difference carries the azimuthal wavenumber information.

\subsubsection{Signal parameters}

\begin{figure}[t]
  \centering
  \includegraphics[width=\figsinglecol]{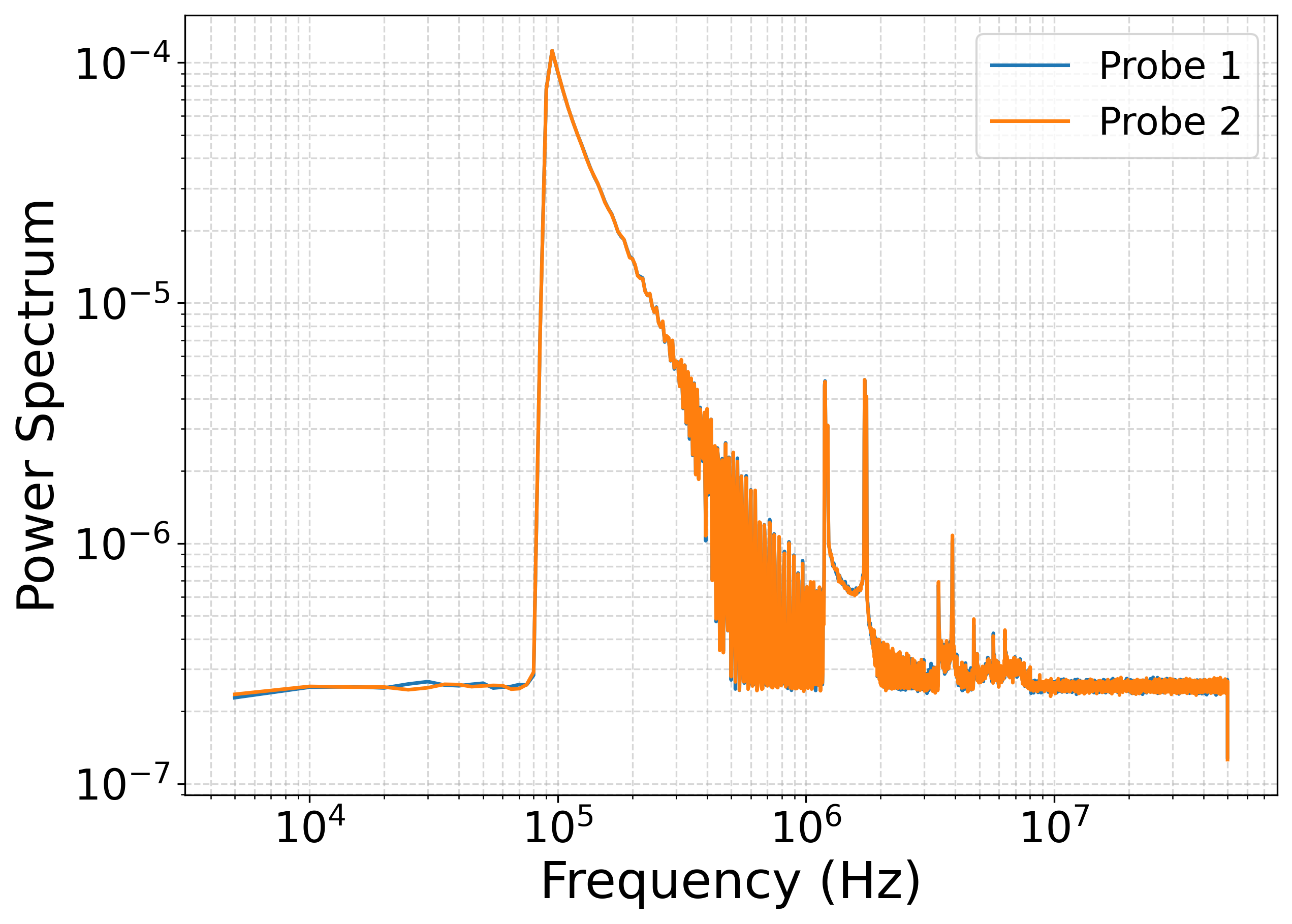}
  \caption{Synthetic-signal power spectrum for the $|r_p|=3.0$~mm, $\alpha=90^\circ$ geometry.
           In real nonlinear saturated experiments, the spectrum would also be affected by mode broadening, energy cascade, and background noise.}
  \label{fig:power_spectrum}
\end{figure}

The amplitude weighting $A_n\propto1/f_n$ yields a power-spectrum scaling $P\propto1/f^2$, capturing the order-of-magnitude high-frequency decay reported in Hall-thruster wave-probe measurements \cite{brown2019,brown2024phd}.
In the implementation, the finite lowest resolved mode frequency prevents a zero-frequency singularity, and the resulting modal amplitudes are normalized before the target SNR is imposed.
With this amplitude choice, the power spectrum is formed by the projection of densely distributed EDI modes onto the frequency axis and appears, under finite frequency resolution, as a quasi-continuous envelope with local peaks (Fig.~\ref{fig:power_spectrum}).

In Fig.~\ref{fig:power_spectrum}, the 200~ms spectrum has a nonmonotonic MHz-scale envelope that rises again after a local dip.
Here a frequency bin denotes the discrete frequency interval obtained from the Fourier transform of a finite-duration time window, with width $\delta f=1/T_\mathrm{seg}$.
When $\delta f$ is large, modes with nearby true frequencies but different $k_z$ values fall within the same bin.

This bin-mixing feature is linked to the multi-order cyclotron resonances and folded structure of EDI.
Unlike nearly monotonic linear dispersions such as IAW, EDI has multiple folded branches near $k_zV_d\approx n\omega_{ce}$, so the same frequency interval can receive contributions from different $k_z$ branches.
Near a folded extremum, the group velocity $v_g=\partial\omega/\partial k_z$ decreases, and nearby wavenumber branches accumulate within a finite frequency bin in a one-dimensional power spectral density (PSD).
The nonmonotonic envelope can therefore be viewed as the projection of the multi-order folded EDI structure onto the frequency axis, rather than as a monotonic power-law turbulent spectrum analogous to IAW.
The absolute spectral amplitude still depends on the modal amplitude envelope, noise floor, frequency-bin width, and segmented spectral-estimation parameters.
Here the spectrum is used as an indication that the synthetic signal retains the multi-order folded EDI features; the physical growth-rate information is obtained from the dispersion solution in Fig.~\ref{fig:dispersion}.

The sampling rate $f_s=100$~MHz is comfortably above the EDI frequency range (highest at $\sim$8~MHz, see Fig.~\ref{fig:dispersion}); the corresponding Nyquist frequency is 50~MHz, about six times the maximum signal frequency, so temporal aliasing is avoided.
The signal-to-noise ratio (SNR), defined here as the ratio of total signal power to total noise power, is set to 10 so that the dominant dispersion structure remains identifiable while finite-noise effects on phase statistics are retained.
The number of modes is about 2000, covering the first five cyclotron resonances and the stable intervals between them.

\subsection{Inversion parameters}

The inversion uses 25 synthetic two-probe baselines to construct multi-geometry phase constraints:
probe spacing is $|r_p|=3.0$--$5.0$~mm with a step of 0.5~mm, and the angle is $\alpha=0^\circ$--$90^\circ$ with a step of 22.5$^\circ$, forming $5\times5$ configurations that span the angular range needed for two-dimensional wave-vector inversion.
The angle $\alpha$ is defined relative to the axial $y$ axis.
For geometry $i$, the two-dimensional baseline components are
$\Delta y^i=|r_p^i|\cos\alpha^i$ and
$\Delta z^i=|r_p^i|\sin\alpha^i$.
Because the prescribed benchmark has $k_y=0$, baselines with small $\Delta z^i$ do not add strong direct sensitivity to $k_z$; instead, they help enforce the absence of axial phase variation and constrain spurious two-dimensional candidates with nonzero $k_y$.

The $k_y$ and $k_z$ search ranges are both $[-1.3k_{z,\max},\,1.3k_{z,\max}]$ with $k_{z,\max}\approx 22$~krad/m, where the $\pm 30\%$ extension beyond the theoretical maximum provides margin for alias-candidate enumeration at the search-domain boundary.
The two-dimensional grid has $N_{k_y}\times N_{k_z}=5001\times5001$ candidate points (not to be confused with the FFT frequency-bin count $N_\mathrm{perseg}$), giving a wave-vector resolution $\Delta k\approx 11$~rad/m.
For each frequency $f$, Eq.~\eqref{eq:joint_loglik} is evaluated on the grid and Eq.~\eqref{eq:mle_kxky} is used to obtain the MLE.
The projected profile in Eq.~\eqref{eq:projected_likelihood} is also saved for $f$--$k_z$ likelihood heat maps and alias-ghost analysis.
The MLE $\hat{k}_z$ is compared with the theoretical azimuthal EDI dispersion, while whether $\hat{k}_y$ is concentrated near zero checks whether the multi-geometry phase constraint converges to the correct azimuthal branch.

The cross spectrum is estimated using segments with $N_\mathrm{perseg}$ samples, so the segment duration is $T_\mathrm{seg}=N_\mathrm{perseg}/f_s$.
The realization count is $L=T_\mathrm{total}/T_\mathrm{seg}$, and the frequency resolution is
$\delta f=1/T_\mathrm{seg}=f_s/N_\mathrm{perseg}$.
With the inversion workflow fixed, the following tests mainly vary $L$ and $\delta f$:
$L$ controls the statistical stability of the cross-spectral phase mean and variance estimates, while $\delta f$ controls the ability of one frequency bin to separate folded dispersion branches.
For a fixed total acquisition time, shortening the segment increases $L$ but coarsens the frequency bin; lengthening the segment reduces $\delta f$ but decreases $L$.

\subsection{Test cases}

The above trade-off motivates two complementary test schemes (Tables~\ref{tab:cases_fixdf}--\ref{tab:cases_fixL}), one varying $L$ at fixed $\delta f$ and the other varying $\delta f$ at fixed $L$, each isolating a distinct error source: suppression of spurious alias peaks by $L$ and local branch separation near folded extrema by $\delta f$.
Both schemes use the same total acquisition times $T=2,\,20,\,50,\,100,\,200$~ms.
The 2~ms lower bound represents an extreme short-sampling limit for fast-insertion measurements with severely limited data.
Fast-insertion probe systems are used in Hall-thruster discharge-channel and near-field diagnostics to reduce probe heating, ablation, and local plasma perturbation associated with long probe dwell \cite{haas1999harp,haas2001,haas1998,lobbia2010}, providing an experimental background for short-dwell wave-probe measurements.
The remaining durations cover the transition from short-duration limitation to statistically sufficient measurements and support the comparison of the fixed-$\delta f$ and fixed-$L$ decoupling paths.

\textbf{Scheme 1: fixed frequency resolution (fixed $\delta f$).}
All durations use $N_\mathrm{perseg}=2000$, corresponding to a segment duration of $T_\mathrm{seg}=20~\mu$s and frequency resolution $\delta f=50$~kHz.
The realization count increases linearly with duration:
\begin{equation}
  L = \frac{T \cdot f_s}{N_\mathrm{perseg}} = \frac{T}{T_\mathrm{seg}} .
  \label{eq:L_from_T}
\end{equation}
This scheme examines the convergence of different resonance orders as $L$ increases from 100 to 10{,}000 at fixed $\delta f$ (Table~\ref{tab:cases_fixdf}).
The 2~ms case has $L=100$ and represents a severely realization-limited condition.

\begin{table}[!tb]
  \centering
  \scriptsize
  \setlength{\tabcolsep}{1.8pt}
  \caption{Test cases for Scheme 1 with fixed $\delta f=50$~kHz and $N_\mathrm{perseg}=2000$.}
  \label{tab:cases_fixdf}
  \begin{tabular}{lllll}
    \toprule
    Duration $T$ & $N_\mathrm{perseg}$ & $\delta f$ & Realizations $L$ & Total samples $N$ \\
    \midrule
    2~ms   & 2{,}000  & 50~kHz  & 100    & $2\times10^5$ \\
    20~ms  & 2{,}000  & 50~kHz  & 1{,}000  & $2\times10^6$ \\
    50~ms  & 2{,}000  & 50~kHz  & 2{,}500  & $5\times10^6$ \\
    100~ms & 2{,}000  & 50~kHz  & 5{,}000  & $1\times10^7$ \\
    200~ms & 2{,}000  & 50~kHz  & 10{,}000 & $2\times10^7$ \\
    \bottomrule
  \end{tabular}
\end{table}

\textbf{Scheme 2: fixed realization count (fixed $L$).}
All durations use $L=100$, and $N_\mathrm{perseg}$ increases proportionally with duration, so the frequency resolution improves as the record becomes longer (Table~\ref{tab:cases_fixL}).
This scheme fixes the statistical sample count at the 2~ms value and reduces $\delta f$ from 50~kHz to 0.5~kHz, thereby isolating the contribution of frequency resolution from the effect of increasing $L$.

The results below use dispersion reconstruction, likelihood heat maps, and local-peak diagnostics to test these two effects.

\begin{table}[!tb]
  \centering
  \scriptsize
  \setlength{\tabcolsep}{1.8pt}
  \caption{Test cases for Scheme 2 with fixed $L=100$ and $N_\mathrm{perseg}$ increasing proportionally with duration.}
  \label{tab:cases_fixL}
  \begin{tabular}{lllll}
    \toprule
    Duration $T$ & $N_\mathrm{perseg}$ & $\delta f$ & Realizations $L$ & Total samples $N$ \\
    \midrule
    2~ms   & 2{,}000   & 50~kHz  & 100 & $2\times10^5$ \\
    20~ms  & 20{,}000  & 5~kHz   & 100 & $2\times10^6$ \\
    50~ms  & 50{,}000  & 2~kHz   & 100 & $5\times10^6$ \\
    100~ms & 100{,}000 & 1~kHz   & 100 & $1\times10^7$ \\
    200~ms & 200{,}000 & 0.5~kHz & 100 & $2\times10^7$ \\
    \bottomrule
  \end{tabular}
\end{table}

\textbf{Shared baseline.} The two schemes coincide at $T=2$~ms ($L=100$, $\delta f=50$~kHz), providing a most-constrained baseline for direct comparison.

\subsection{Error metric}

For each mode $n$ with true azimuthal wavenumber $k_{z,n}$, the relative error is
\begin{equation}
  \epsilon_\mathrm{rel}(n) =
  \frac{|\hat{k}_z(f_n) - k_{z,n}|}{|k_{z,n}| + \epsilon_0} \times 100\% ,
  \label{eq:err_rel}
\end{equation}
where $\epsilon_0=10^{-10}$~rad/m prevents division by zero.
The median over all modes in a frequency band is used as the overall performance metric, where the frequency bands correspond to the five cyclotron-resonance ranges listed in Table~\ref{tab:threshold}.
A 10\% threshold is adopted for convergence, consistent with the convergence criterion used in the IAW benchmark of Liu and Jorns~\cite{liu2025}.

Near folded extrema, the true $k_{z,n}$ can become locally small, so isolated relative-error spikes may be amplified by the denominator.
Such spikes should therefore be interpreted together with the dispersion scatter plots.

\section{Results}
\label{sec:results}

\subsection{Beall baseline: confirmation of aliasing}

\begin{figure}[t]
  \centering
  \includegraphics[width=\figsinglecol]{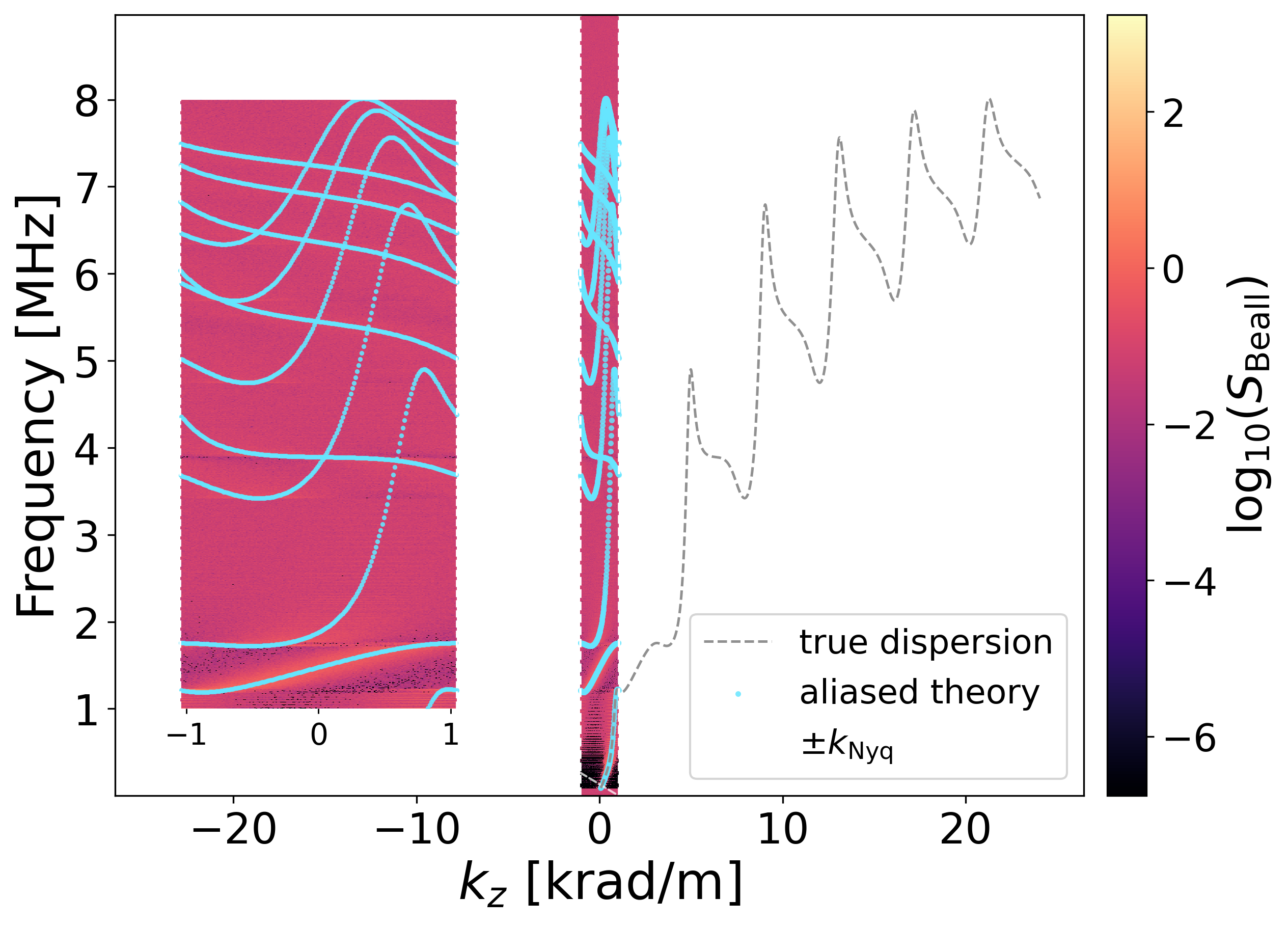}
  \caption{Aliasing heat map from conventional Beall analysis of the EDI signal.
           Color: power-weighted Beall histogram.
           Light gray dashed lines: true theoretical EDI dispersion; cyan points: theoretical dispersion folded into the Beall-visible interval; white vertical dashed lines: $\pm k_\mathrm{Nyq}\approx\pm1.05$~krad/m.
           The left inset zooms into $k_z\in[-1.12,1.12]$~krad/m and $f\in[1.0,8.0]$~MHz, showing that high-wavenumber branches are compressed into a narrow aliased band.}
  \label{fig:beall}
\end{figure}

Figure~\ref{fig:beall} shows conventional Beall analysis applied to the EDI signal for the $|r_p|=3.0$~mm, $\alpha=90^\circ$ baseline.
This baseline has the smallest probe spacing among the 25 configurations, giving the largest $k_\mathrm{Nyq}$ and therefore the most favorable Beall observable range.
Color represents the power-weighted Beall histogram (Sec.~II.B) inferred from the two-probe phase difference.
Because the phase difference only gives the folded wavenumber within $(-k_\mathrm{Nyq},k_\mathrm{Nyq}]$, the heat map is restricted to the Nyquist interval
$\pm k_\mathrm{Nyq}\approx\pm1.05$~krad/m, whereas the true EDI dispersion extends to about 22~krad/m.
The inset shows the central Nyquist interval, where several high-wavenumber branches are folded into the same narrow wavenumber window.
This confirms the fundamental Nyquist constraint of Beall analysis: the cross-spectral phase determines only the phase difference modulo $2\pi$, so branches with wavelengths comparable to or smaller than the probe spacing are folded back into the Nyquist interval~\cite{beall1982,liu2025,liu2025iepc}.
The remainder of this section compares the spatial anti-aliasing inversion against this Beall baseline.

\subsection{Evolution of dispersion reconstruction with data duration}

\begin{figure*}[t]
  \centering
  \captionsetup{skip=1pt}
  \captionsetup[subfigure]{font=footnotesize,skip=2pt}
  \begin{subfigure}[t]{\figthreeperrow}
    \centering
    \includegraphics[width=\linewidth]{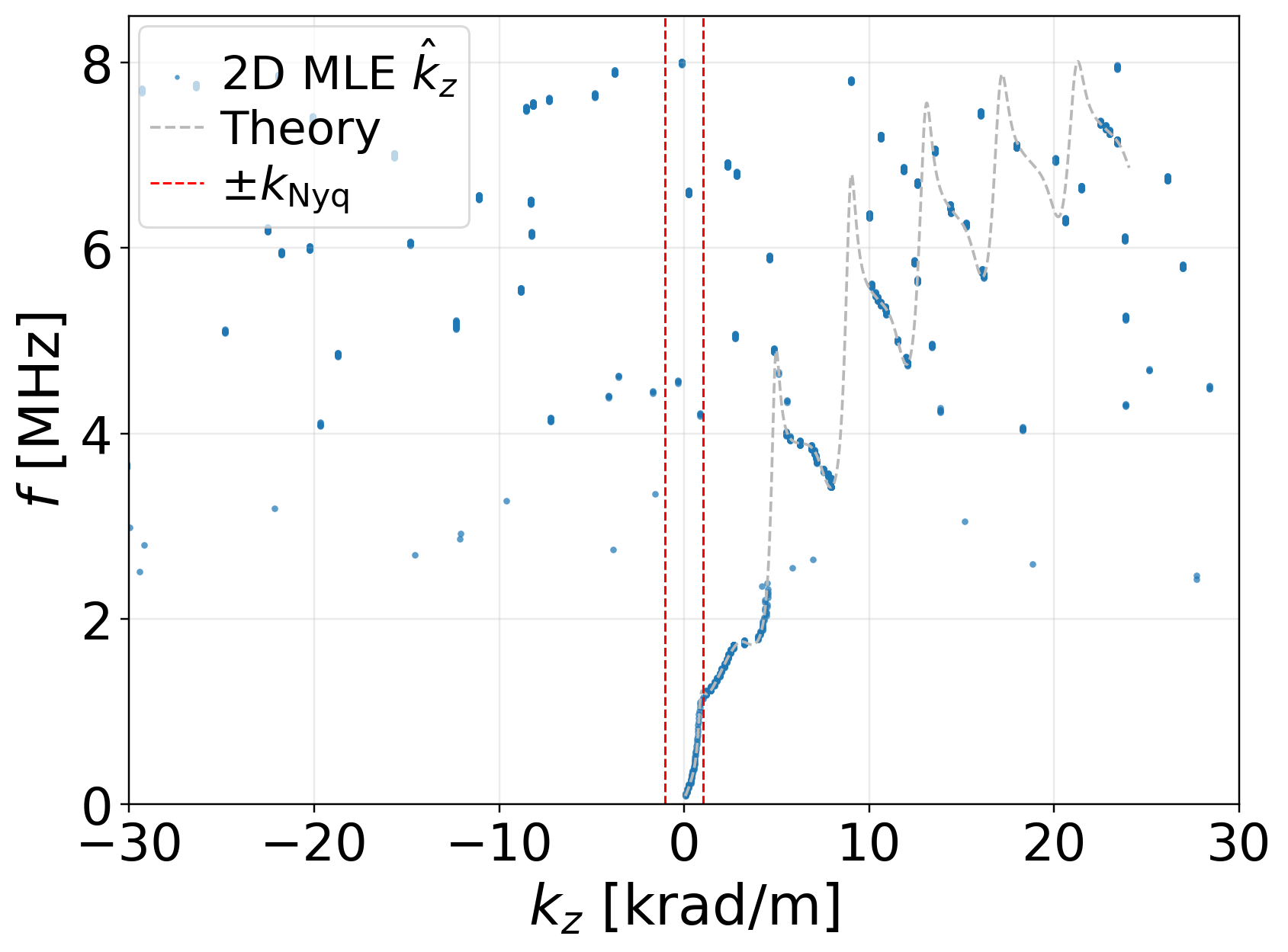}
    \caption{}
  \end{subfigure}
  \hfill
  \begin{subfigure}[t]{\figthreeperrow}
    \centering
    \includegraphics[width=\linewidth]{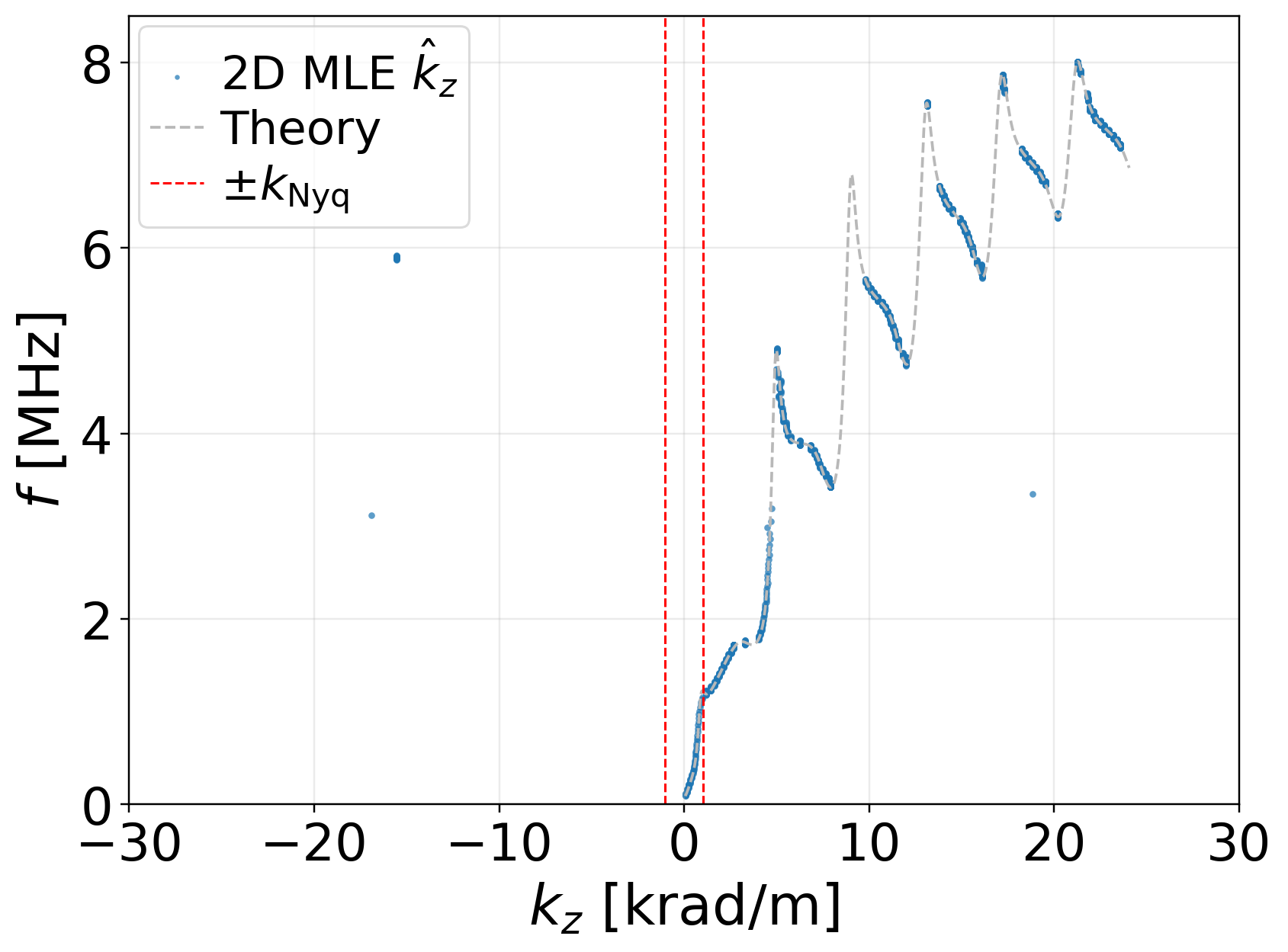}
    \caption{}
  \end{subfigure}
  \hfill
  \begin{subfigure}[t]{\figthreeperrow}
    \centering
    \includegraphics[width=\linewidth]{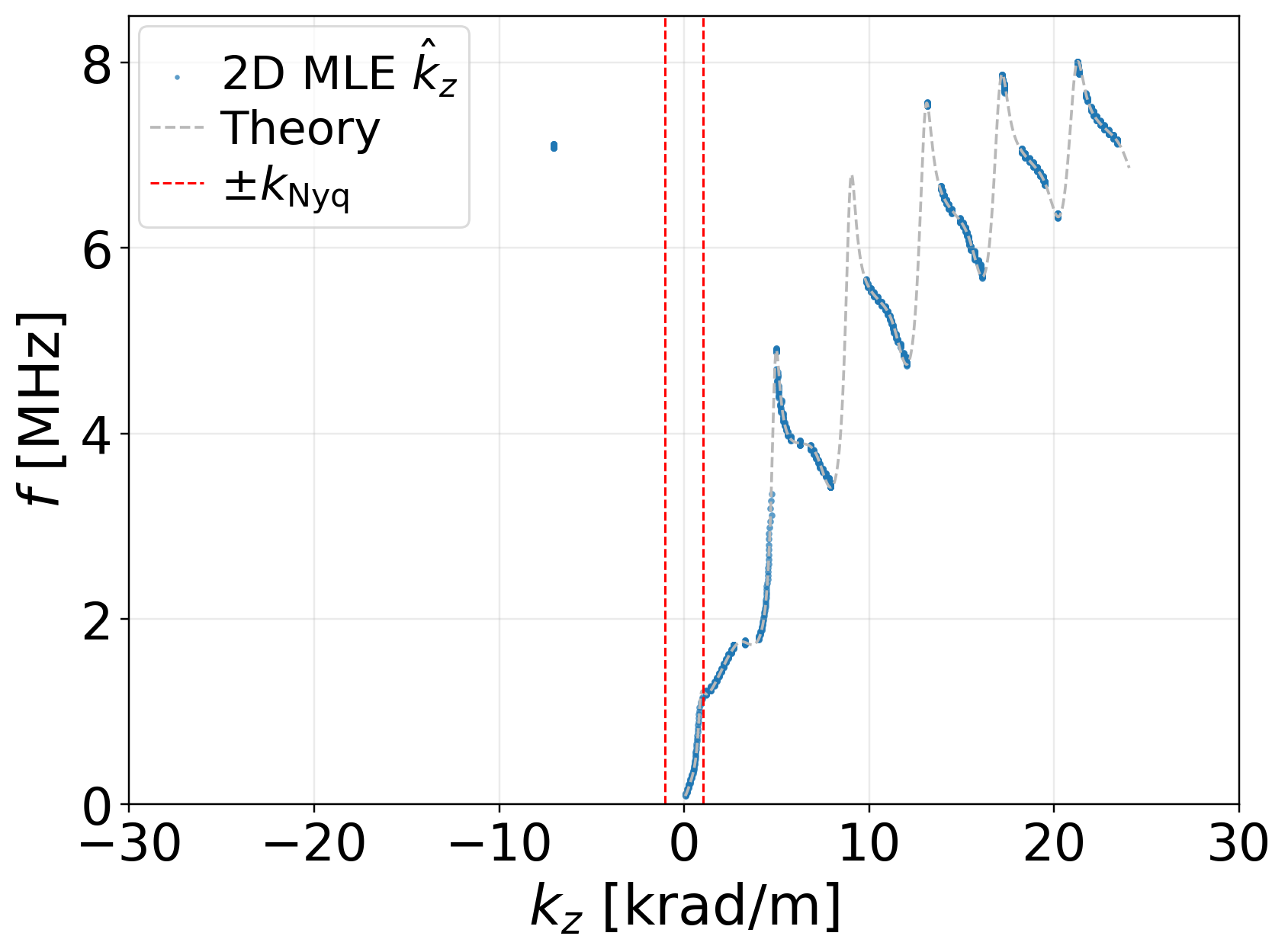}
    \caption{}
  \end{subfigure}
  \\[-2pt]
  \begin{subfigure}[t]{\figthreeperrow}
    \centering
    \includegraphics[width=\linewidth]{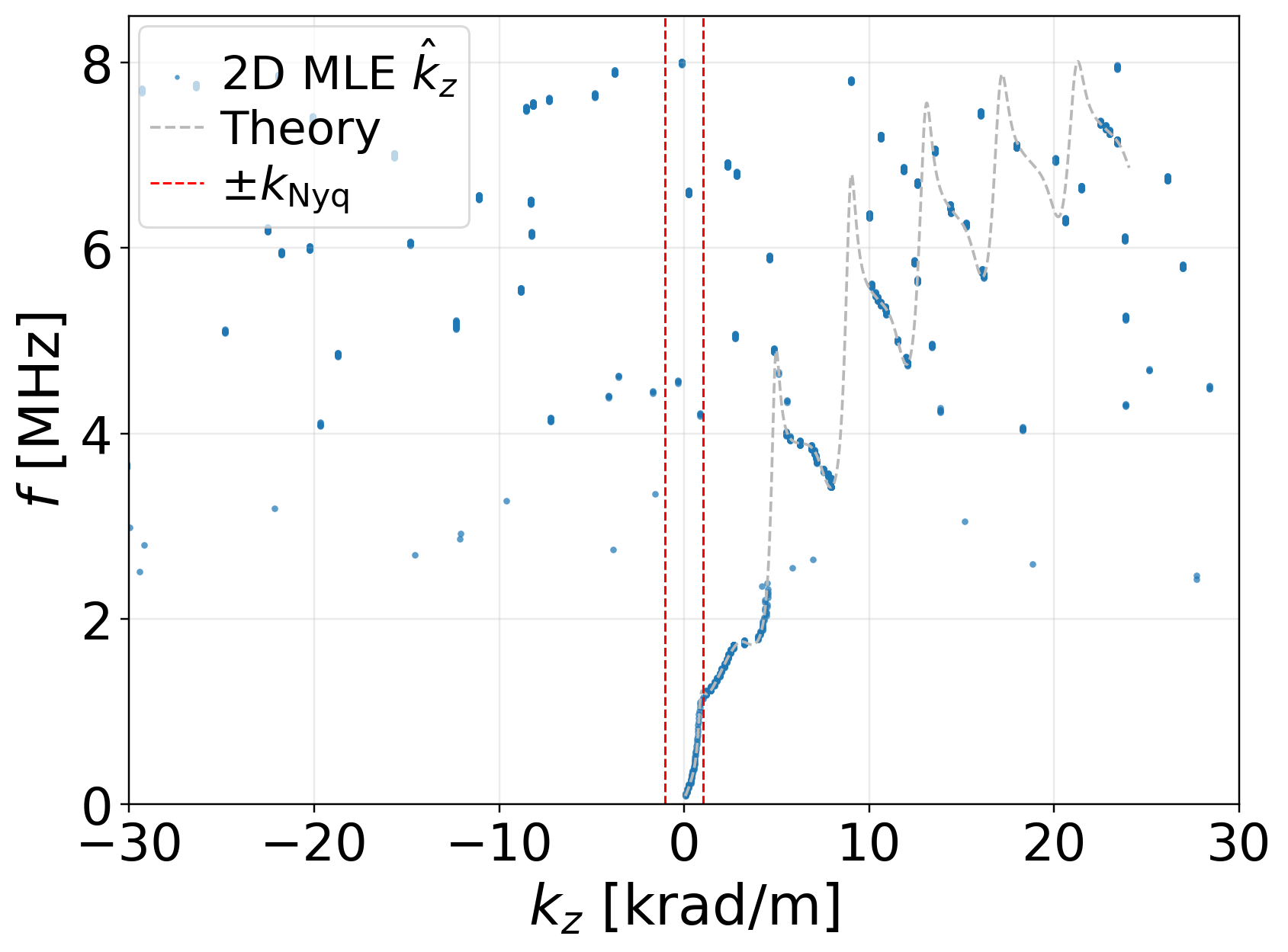}
    \caption{}
  \end{subfigure}
  \hfill
  \begin{subfigure}[t]{\figthreeperrow}
    \centering
    \includegraphics[width=\linewidth]{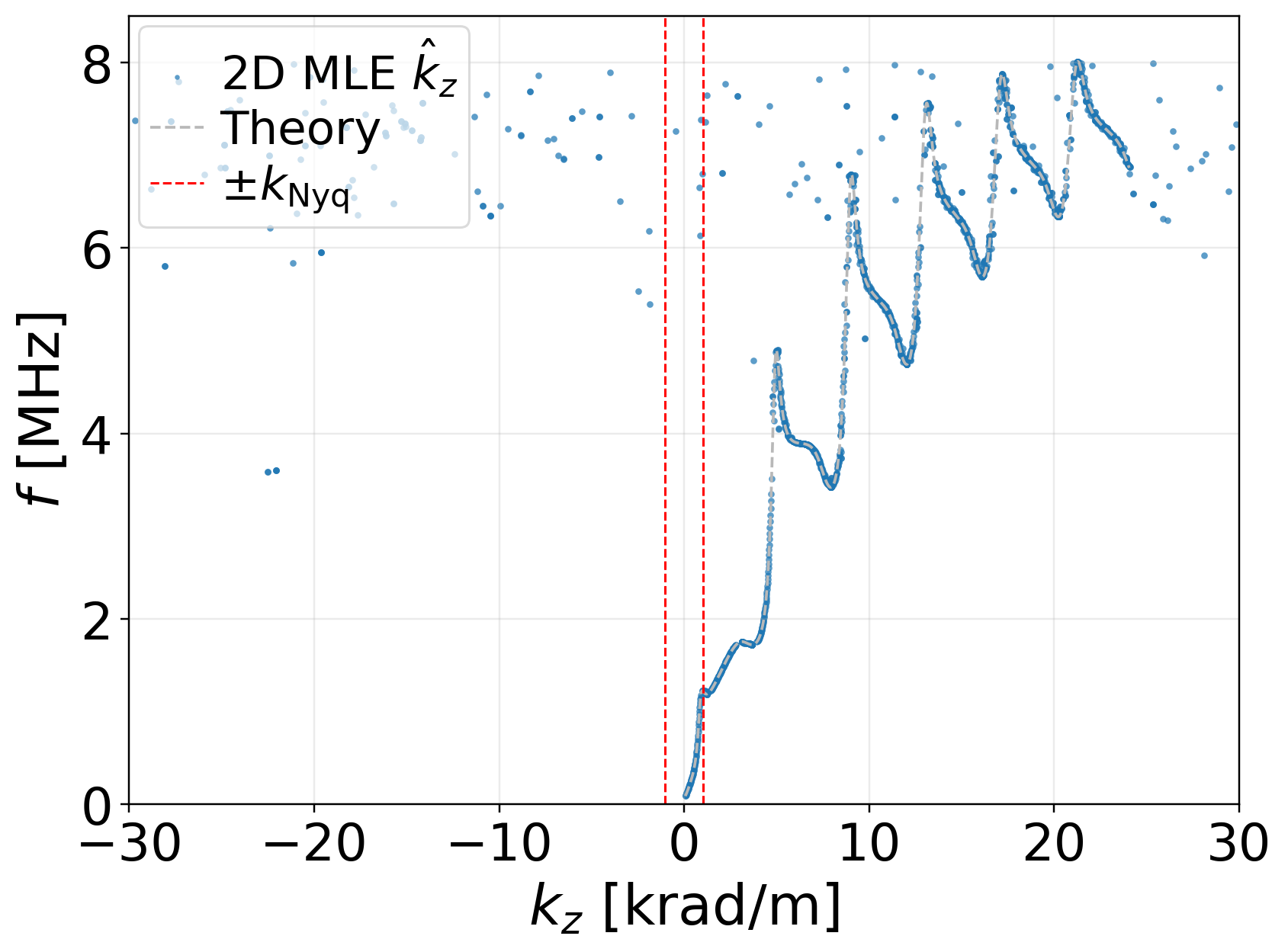}
    \caption{}
  \end{subfigure}
  \hfill
  \begin{subfigure}[t]{\figthreeperrow}
    \centering
    \includegraphics[width=\linewidth]{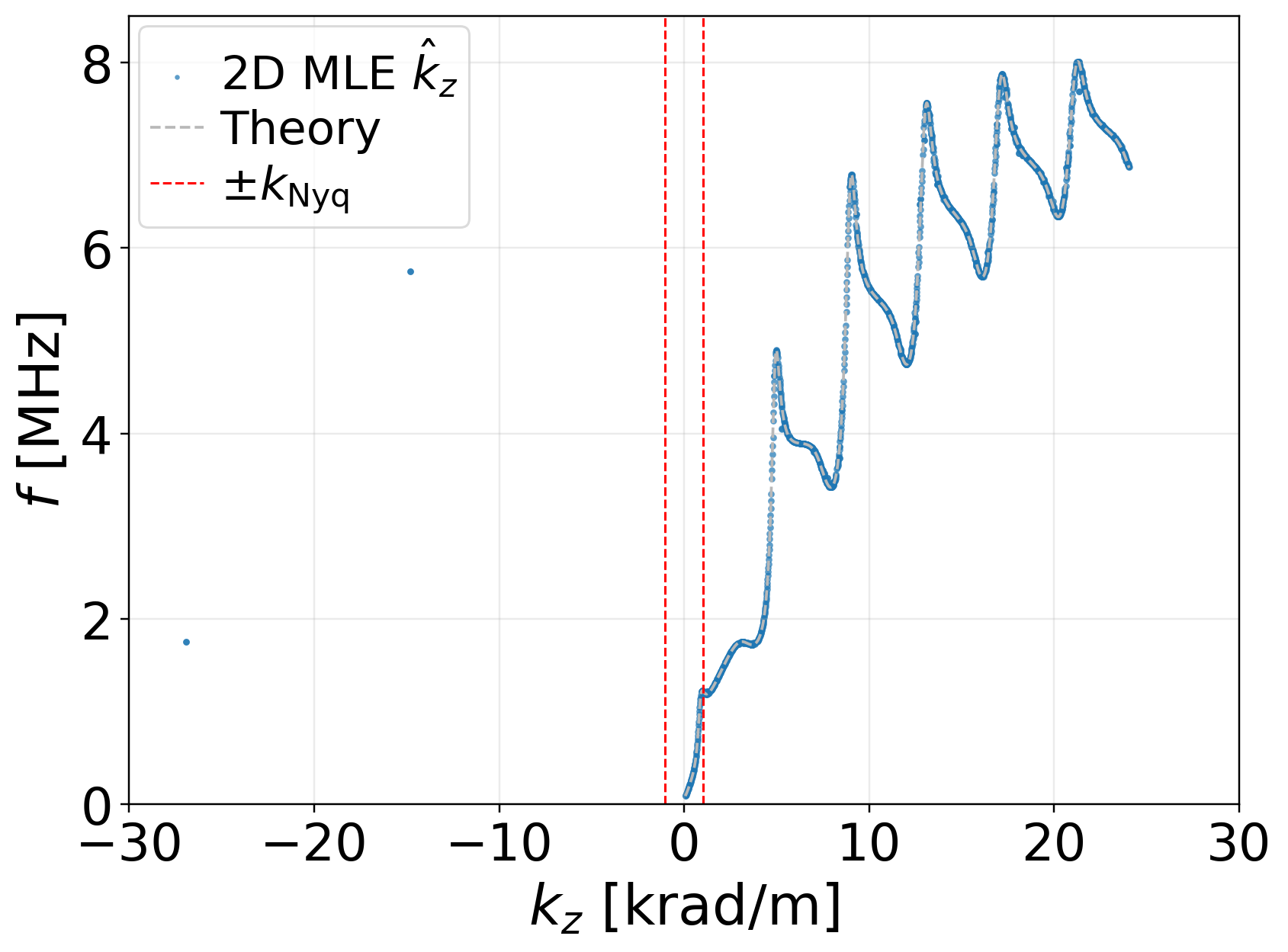}
    \caption{}
  \end{subfigure}
  \caption{EDI dispersion reconstruction for the two schemes at three representative durations.
           Panels (a--c) show Scheme 1 at 2, 50, and 200~ms; panels (d--f) show Scheme 2 at 2, 50, and 200~ms.
           Blue points: $k_z$ projections of the two-dimensional MLE; light gray dashed lines: theoretical dispersion; red vertical dashed lines: $\pm k_\mathrm{Nyq}$.
           The 2~ms case is the shared most-constrained baseline; as duration increases, the two paths separately show the suppression of spurious peaks by increasing $L$ and the compression of folded-region local errors by decreasing $\delta f$.}
  \label{fig:scatter_both}
\end{figure*}

Figure~\ref{fig:scatter_both} compares the reconstructed dispersion for the two schemes at three representative durations.
The top row is Scheme 1, where $\delta f=50$~kHz is fixed and $L$ increases with duration.
The bottom row is Scheme 2, where $L=100$ is fixed and $\delta f$ decreases with duration.
The three columns correspond to 2, 50, and 200~ms.

At 2~ms, the two schemes coincide at $L=100$, $\delta f=50$~kHz: the low-order resonances ($n=1,2$) have already converged, whereas high-order resonance points are randomly distributed.
As the duration grows, both paths drive high-order resonances toward convergence, but through different mechanisms.

In Scheme 1, the mechanism is statistical: increasing $L$ from 100 to 10{,}000 strongly reduces random outliers and spurious alias peaks.
As $L$ grows, the third to fifth resonances cross the 10\% median-error threshold by 20~ms (Fig.~\ref{fig:heatmap_both}); from 50 to 200~ms the dominant dispersion branches stabilize further, with the single-peak accessible wavenumber reaching about 22~krad/m at 200~ms.

The scatter points in Fig.~\ref{fig:scatter_both} are single-peak MLE outputs --- for each frequency bin, only the maximum-likelihood $(k_y,k_z)$ candidate is retained.
Under Scheme 1's coarse frequency bins, multiple folded branches compete for that single peak, so the scatter mainly displays dominant branches rather than the full multi-valued structure.

In Scheme 2, the mechanism is spectral: $\delta f$ decreases from 50~kHz to 0.5~kHz while $L$ remains 100.
At 20~ms ($\delta f=5$~kHz), the errors of $n=4,5$ decrease substantially while remaining above the threshold.
At 50~ms ($\delta f=2$~kHz), all five resonances meet the 10\% median-error convergence criterion, and the scatter near folded points is more concentrated than in Scheme 1 at the same duration.

The more complete appearance of folded multi-valued structures at 50--200~ms should be interpreted through the bin-by-bin nature of the estimator.
Reducing $\delta f$ gives denser frequency-axis sampling, so branches that were mixed within coarse bins or secondary in single-peak selection have more opportunities to become local dominant peaks at adjacent fine frequency points.
Thus Scheme 2 improves the visibility of bin-by-bin single-peak recovery, whereas strict same-frequency multi-peak inversion requires an additional multimodal selection step.
At short durations, the small $L=100$ still leaves the result vulnerable to random outliers and spurious peaks.

\subsection{Likelihood heat maps and alias ghosts}

\begin{figure*}[t]
  \centering
  \begin{subfigure}[t]{\figtwoperrow}
    \centering
    \includegraphics[width=\linewidth]{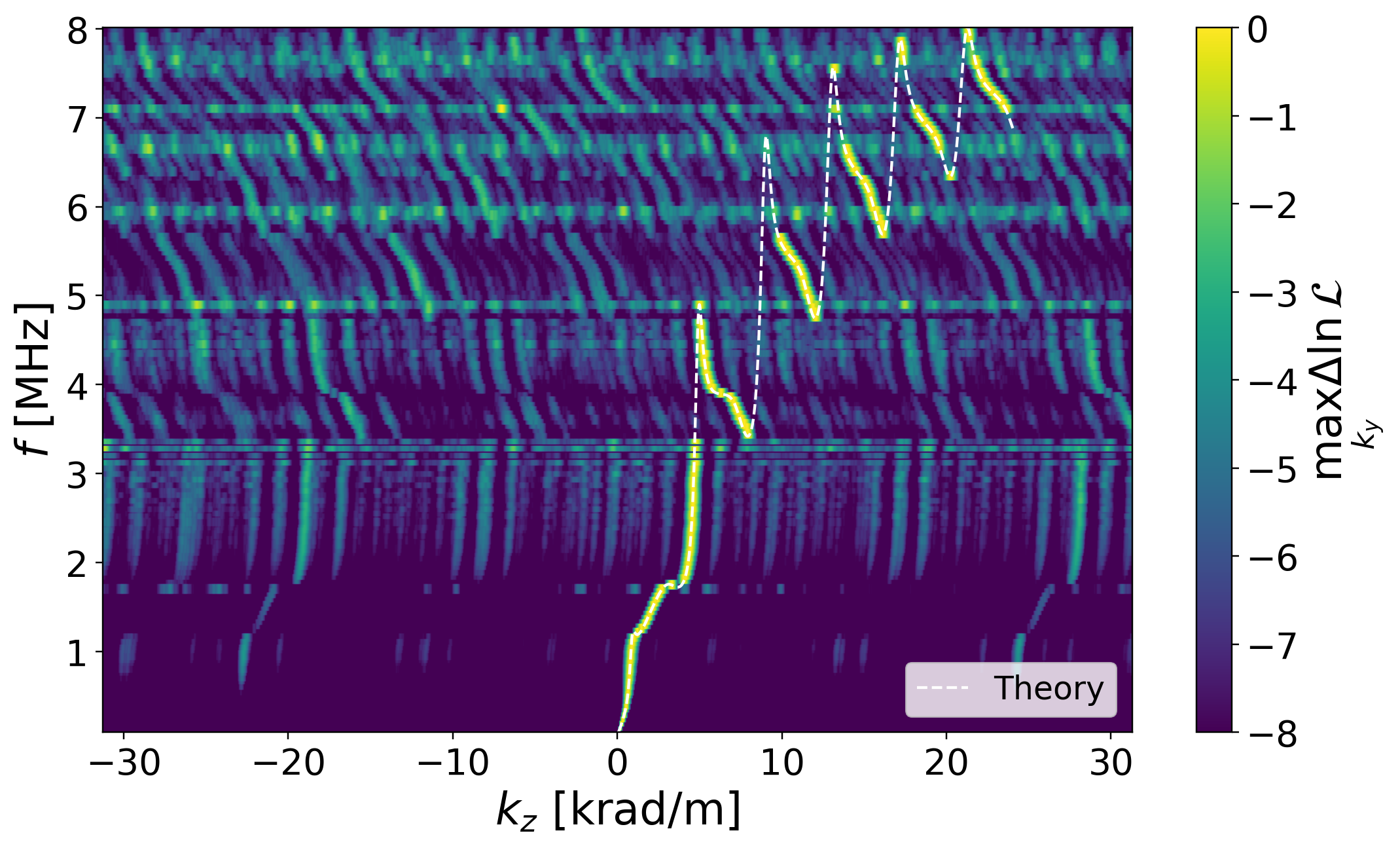}
    \caption{}
  \end{subfigure}
  \begin{subfigure}[t]{\figtwoperrow}
    \centering
    \includegraphics[width=\linewidth]{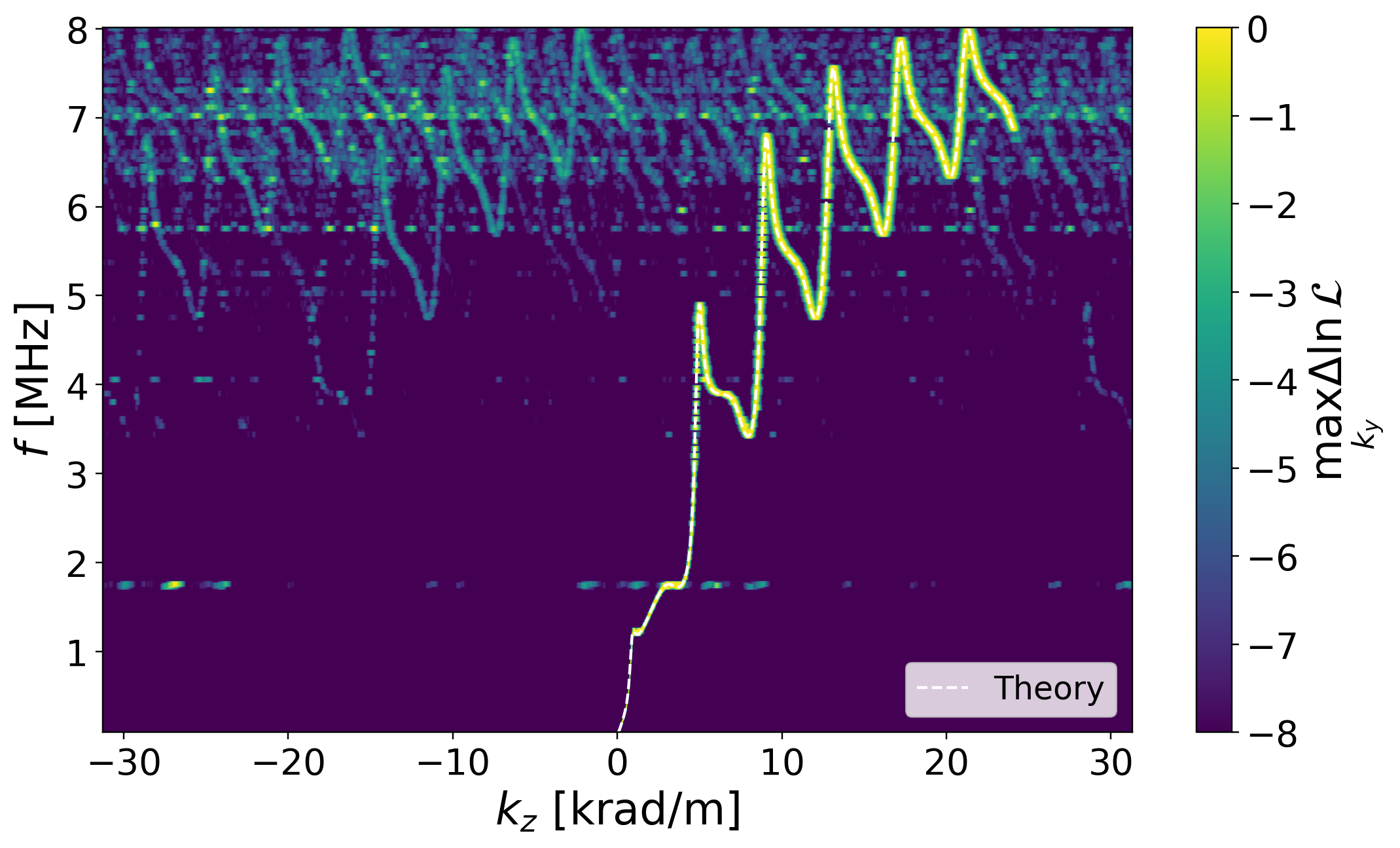}
    \caption{}
  \end{subfigure}
  \caption{Comparison of two-dimensional projected likelihood heat maps at 200~ms.
           Panel (a): Scheme 1 with $L=10{,}000$ and $\delta f=50$~kHz.
           Panel (b): Scheme 2 with $L=100$ and $\delta f=0.5$~kHz.
           Color: relative log likelihood after maximizing over $k_y$, $\max_{k_y}\Delta\ln\mathcal{L}(k_y,k_z)$; white dashed lines: theoretical dispersion.}
  \label{fig:heatmap_compare}
\end{figure*}

The following diagnostics decompose the inversion errors caused by folded EDI dispersion into three levels.
First, Fig.~\ref{fig:heatmap_compare} shows the global topology of the two-dimensional joint likelihood projected onto the $f$--$k_z$ plane, identifying the true dispersion ridge, alias ghosts, and local broadening near folded points.
Second, Fig.~\ref{fig:likelihood_metrics} converts this topology into two scalar diagnostics: the likelihood gap between the true and competing peaks and the main-peak width.
These metrics separate global peak-competition failure from local folded-region broadening.
Third, Figs.~\ref{fig:topn_likelihood}--\ref{fig:bin_mixing_ablation} examine secondary local peaks beyond the single-peak output and distinguish whether missing branches are discarded by single-peak selection, pushed to lower rank by competing peaks, or fail to form an independent local peak in the current frequency bin.
This sequence is used to connect the visual reconstruction quality to the underlying likelihood topology.

\textbf{(1) Ridge sharpening.}
In Scheme 1, $L=10{,}000$ and $\delta f=50$~kHz produce sharp likelihood ridges, showing that many realizations stabilize the joint likelihood.
Some broadening remains near the first-resonance folded points, consistent with possible multi-branch contributions within wide frequency bins.
In Scheme 2, $L=100$ and $\delta f=0.5$~kHz also give clear ridges, and the broadening near folded points is weaker, suggesting that finer $\delta f$ can compress local phase mixing near folded extrema.

\textbf{(2) Suppression of alias ghosts.}
Mirror-like ghosts on the negative-$k_z$ side arise from the $2\pi$ periodicity of phase differences.
When probe spacings have approximate common divisors, corresponding spurious peaks can partially overlap and form residual secondary maxima.
Both schemes suppress these ghosts substantially: Scheme 1 relies on phase averaging over many realizations, whereas Scheme 2 may benefit from more localized frequency sampling under fine $\delta f$, which helps focus the cross-geometry likelihood.
At the dominant-ridge level in the 200~ms cases, the ghost intensity in both paths is lower than that of the true dispersion, and single-peak MLE shows no systematic misidentification.
However, alias candidates can still remain among secondary local peaks, as further separated by the top-$N$ diagnostics below.

Together, Fig.~\ref{fig:heatmap_compare} indicates that $L$ and $\delta f$ are linked to different error types: $L$ is more directly associated with stable suppression of spurious peaks by the joint likelihood, whereas $\delta f$ is more directly associated with the likelihood-ridge width near folded extrema.

\subsection{Quantifying likelihood-peak competition}

The heat maps show the overall likelihood-ridge shape, but understanding the convergence behavior of high-order resonances requires quantifying the relative height of the true and competing peaks.
The following metrics are defined on the projected likelihood profile $\ln\mathcal{L}^{\rm proj}_f(k_z)$ from Eq.~\eqref{eq:projected_likelihood}.
For each frequency $f$, we define the log-likelihood gap between the true peak and the strongest non-true candidate as

\begin{equation}
  \Delta\ln \mathcal{L}_f =
  \ln \mathcal{L}^{\rm proj}_f(k_{z,\mathrm{true}}) -
  \max_{k_z\in\mathcal{A}_f}\ln \mathcal{L}^{\rm proj}_f(k_z),
  \label{eq:likelihood_gap}
\end{equation}

where $\mathcal{A}_f$ is the projected $k_z$ candidate region after excluding the neighborhood of the true branch.
The neighborhood is defined as $|k_z-k_{z,\mathrm{true}}|<\Delta k_\mathrm{tol}$, with
$\Delta k_\mathrm{tol}=\max(0.5~\mathrm{krad/m},0.05|k_{z,\mathrm{true}}|)$, chosen to exclude the immediate vicinity of the true peak (well above the wave-vector grid step $\Delta k\approx 11$~rad/m) while leaving the rest of the projected wave-vector domain for competition.
Thus $\Delta\ln\mathcal{L}_f>0$ means that the true branch has a higher likelihood than the strongest competing peak.
If $\Delta\ln\mathcal{L}_f<0$, single-peak MLE is more likely to lock onto a spurious alias peak or a same-frequency competing branch.

Local uncertainty in the folded region is characterized by the main-peak width.
Let the main peak at a given frequency be located at $k_\mathrm{max}$.
The two boundaries $k_-$ and $k_+$ are defined at a log-likelihood drop $\Delta_c$:

\begin{equation}
  \begin{aligned}
  W_f &= k_+ - k_- ,\\
  \ln\mathcal{L}^{\rm proj}_f(k_\pm)
  &= \ln\mathcal{L}^{\rm proj}_f(k_\mathrm{max})
  - \Delta_c .
  \end{aligned}
  \label{eq:likelihood_width}
\end{equation}

We use $\Delta_c=2$, corresponding to a likelihood ratio $e^{-2}\approx 0.135$ and therefore to a conservative finite-width contour around the dominant peak, and approximate the contour boundaries with neighboring discrete grid points.
A larger $W_f$ means weaker localization of the dominant peak in $k_z$.
Near folded extrema, this width is also affected by nearby branch mixing and peak switching.

\begin{figure*}[t]
  \centering
  \includegraphics[width=\figwideone]{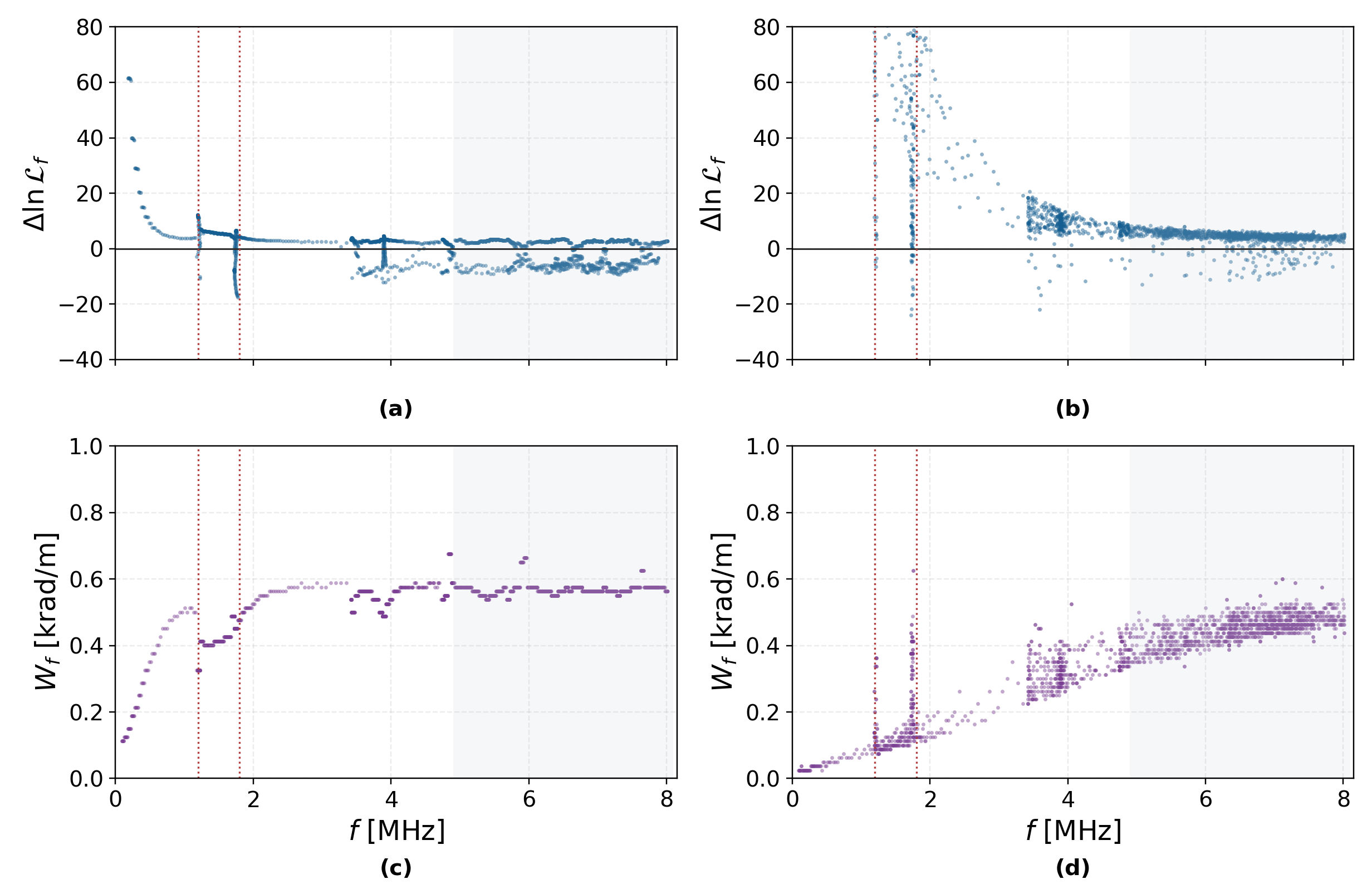}
  \caption{Likelihood-peak competition metrics for the two decoupling schemes at 200~ms.
           (a,b) Log-likelihood gap $\Delta\ln\mathcal{L}_f$ between the true branch and the strongest non-true candidate.
           (c,d) Main-peak width $W_f$.
           Panels (a,c): Scheme 1; panels (b,d): Scheme 2.
           Red vertical dashed lines mark the first-resonance folded extrema near 1.2 and 1.8~MHz; gray shading: higher-order resonance band.}
  \label{fig:likelihood_metrics}
\end{figure*}

Figure~\ref{fig:likelihood_metrics} quantifies the differences observed in Fig.~\ref{fig:heatmap_compare}.
Scheme 1 already recovers the dominant dispersion ridge at 200~ms, but the high-frequency band still contains a group of points with $\Delta\ln\mathcal{L}_f<0$.
Scheme 2, at the same total duration, retains finer frequency sampling; the positive gap of the true peak over competitors becomes more continuous and $W_f$ is clearly reduced.
Local jumps in both width and gap still appear near the two folded extrema.
Beyond confirming the qualitative $L$/$\delta f$ attribution of Fig.~\ref{fig:heatmap_compare}, the metrics reveal that local jumps in both $\Delta\ln\mathcal{L}_f$ and $W_f$ near the folded extrema persist even when $L=10{,}000$; this indicates that the folded region involves branch mixing and peak competition beyond statistical sample-mean noise and cannot be removed by increasing $L$ alone.

\subsection{Multi-valued branches under single-peak MLE}

The previous metrics use single-peak MLE, which returns one peak at each frequency.
To test whether missing branches are absent from the joint likelihood, we further extract the top three local maxima from the projected two-dimensional likelihood profile
$\max_{k_y}\ln\mathcal{L}_f(k_y,k_z)$ at each frequency.
Local peaks are required to be separated by at least 0.5~krad/m to avoid counting the same broad peak multiple times on the discrete grid.

\begin{figure*}[t]
  \centering
  \includegraphics[width=\figwideone]{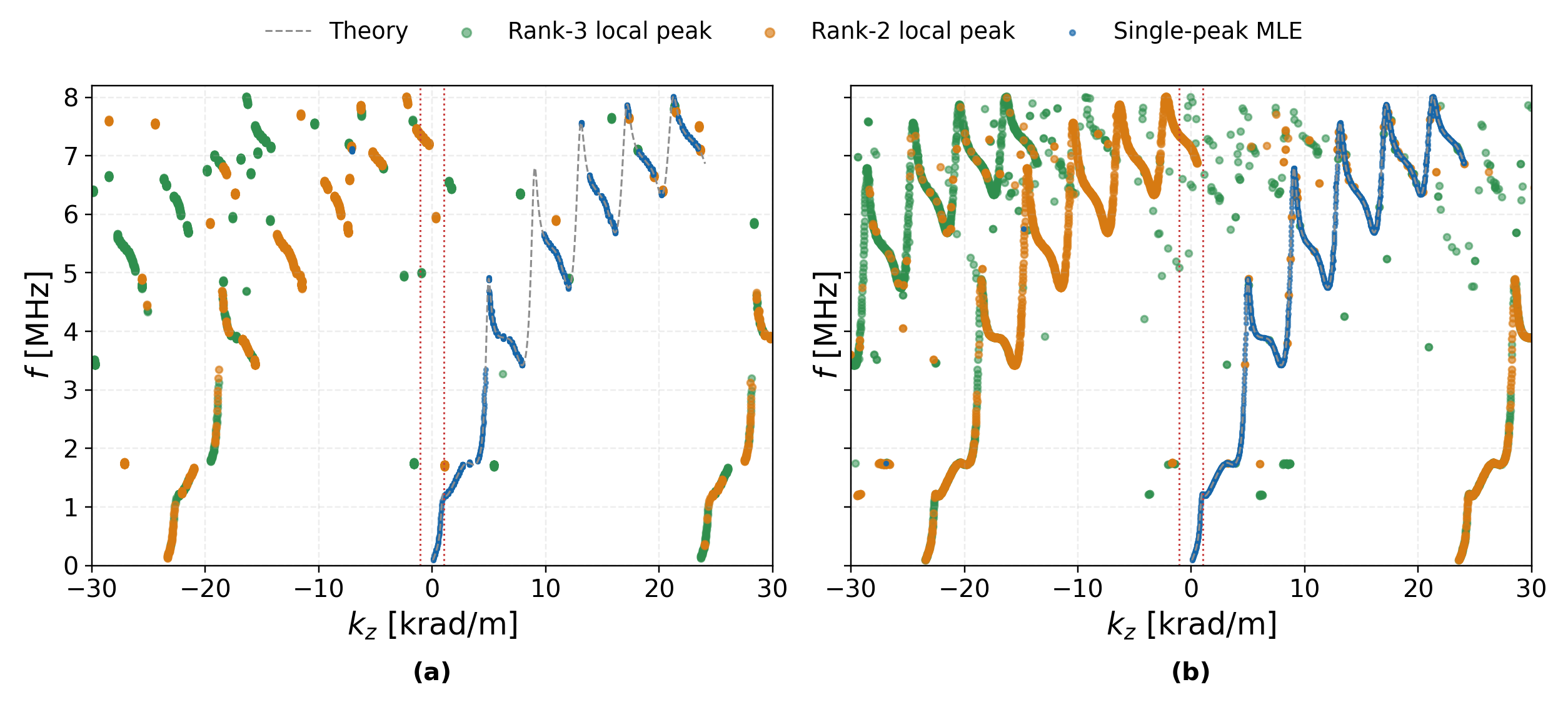}
  \caption{Comparison of single-peak output and top-$N$ local likelihood-peak candidates at 200~ms.
           Gray dashed lines: theoretical EDI dispersion; blue points: $k_z$ projections of single-peak MLE; orange and green open circles: second and third local likelihood-peak candidates; red vertical dashed lines: $\pm k_\mathrm{Nyq}$.
           The top-$N$ candidates include both true folded-branch peaks and alias candidates produced by phase wrapping, before any branch-continuity or physics-based filtering.}
  \label{fig:topn_likelihood}
\end{figure*}

Figure~\ref{fig:topn_likelihood} shows that single-peak MLE outputs the dominant branch in each frequency bin rather than all feasible $k_z$ branches at that frequency.
Some theoretical folded branches missing from the single-peak scatter plot still appear as the second or third local likelihood peak, indicating that the multi-geometry phase constraints retain part of the secondary branch information.

Consider the right folded point of the first resonance near 1.8~MHz.
In Scheme 1, $\delta f=50$~kHz, so one frequency bin covers approximately 1.775--1.825~MHz.
In Scheme 2, $\delta f=0.5$~kHz, so the same 50~kHz interval is resolved into about 100 fine bins.
Because folded regions satisfy the multi-valued mapping
$f\rightarrow\{k_{z,1},k_{z,2},\ldots\}$, the coarse bin in Scheme 1 is more likely to contain multiple $k_z$ branches on both sides of the folded point.
One branch may become the blue main peak because of higher power or stronger cross-geometry phase consistency, while another branch that still lies on the theoretical dispersion may appear as an orange or green secondary local peak.
By reducing bin width and increasing frequency sampling, Scheme 2 allows these branches, which compete within a coarse bin, to appear as main peaks or local candidate peaks at neighboring fine frequency points.
Thus the visibility of multi-valued structures is better in Scheme 2.

Secondary candidates near the theoretical dispersion in Fig.~\ref{fig:topn_likelihood} without corresponding blue single-peak points reflect the output difference caused by bin-by-bin main-peak selection.
These differ from missing-candidate gaps, where the theoretical branch has no representation in the top-$N$ output.
In the latter case, under the current discrete frequency bins and top-3 local-peak truncation, the branch lacks a sufficiently prominent independent candidate peak.
Possible causes include complex-vector mixing of multiple branches in a coarse frequency bin, sparse frequency sampling missing a continuous theoretical branch, and alias or other competing peaks outranking the true branch in the local likelihood ordering.
The mechanism behind these gaps is separated with the rank diagnostic and ablation test below.

\begin{figure*}[t]
  \centering
  \includegraphics[width=\figwideone]{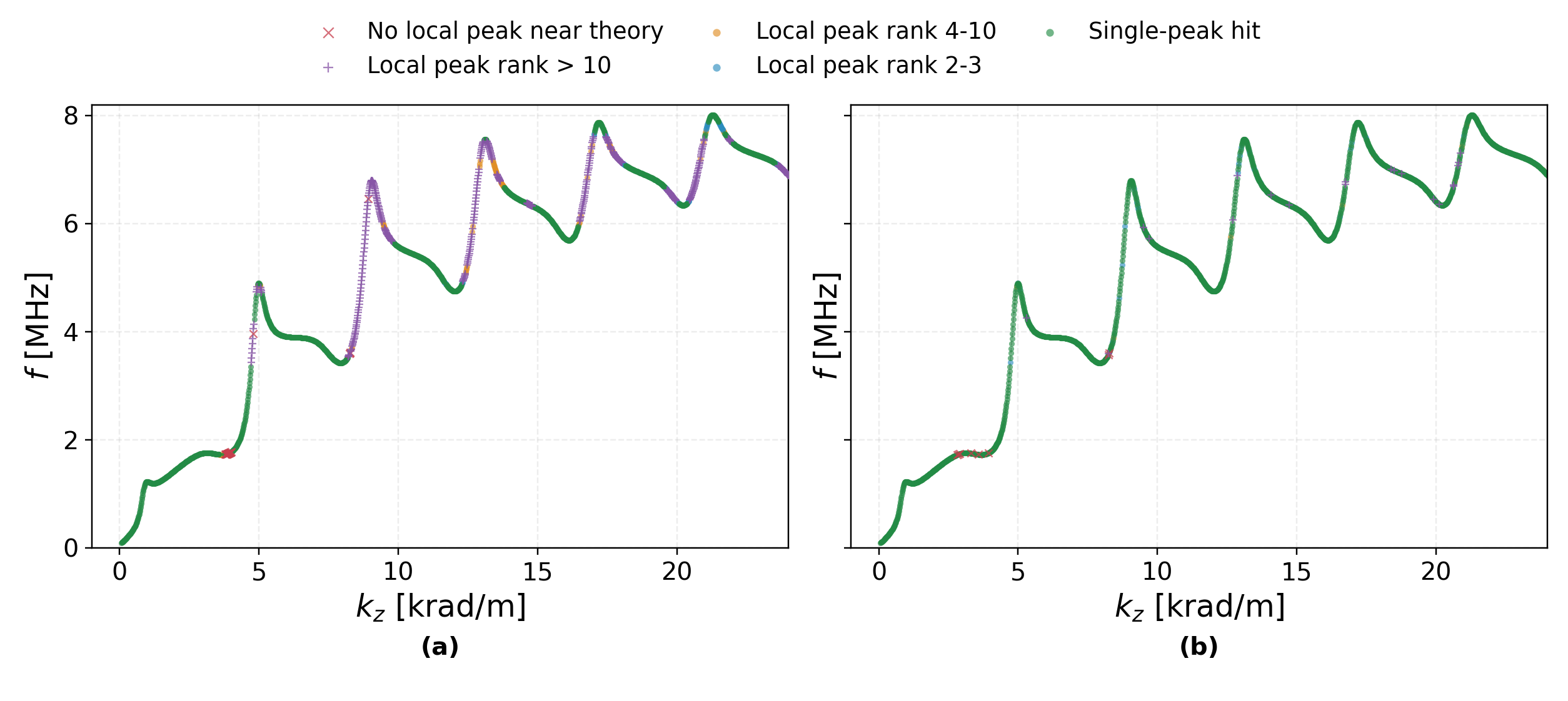}
  \caption{Local likelihood-peak rank diagnostics in the neighborhood of theoretical modes at 200~ms.
           Colors indicate the first local-peak rank that hits the neighborhood of each theoretical mode in the projected two-dimensional likelihood profile.
           Red crosses mark theoretical neighborhoods where no identifiable local peak is formed under the present criterion (distinct from candidates retained in the likelihood function but truncated by top-$N$).}
  \label{fig:branch_gap_diagnostics}
\end{figure*}

To further distinguish these gaps, we use the 2000 theoretical modes injected into the present simulation as targets and examine, point by point, the local-peak rank in the neighborhood of each theoretical $k_z$.
Figure~\ref{fig:branch_gap_diagnostics} classifies theoretical modes into five categories:
rank-1 points hit by single-peak MLE, points hit by the second or third candidate peak, points appearing only at rank 4--10 or beyond rank 10, and points for which no identifiable local peak forms in the theoretical neighborhood.

In Scheme 1, single-peak MLE hits 69.7\% of theoretical modes directly and the second/third candidates add 3.8\%, leaving 4.4\% at rank 4--10, 21.0\% beyond rank 10, and 1.2\% with no identifiable local peak.
In Scheme 2, single-peak coverage rises to 94.8\% and cumulative top-3 coverage reaches 97.4\%, with the no-local-peak fraction dropping to 0.4\%.
This jump tracks the frequency-bin occupancy: Scheme 1's 2000 modes occupy only 156 FFT bins (98.9\% of modes share a bin), whereas Scheme 2 spreads them across 1813 bins (17.6\% share a bin).
The main gaps in Scheme 1 therefore arise because local candidate peaks in coarse bins are pushed to lower ranks by alias peaks or other competitors, rather than from complete absence of a local peak in the theoretical neighborhood.
Finer $\delta f$ reduces same-bin mode competition and densifies frequency sampling, greatly improving the coverage of theoretical branches by main peaks and low-rank candidates.

To test whether same-bin multi-branch competition can change the identity of the main peak, we perform a focused ablation on a gap point near the first folded resonance in Scheme 1.
The target mode is at $f=1.727$~MHz and $k_z=3.782$~krad/m.
Its nearest FFT bin center is 1.750~MHz, and this 50~kHz bin contains 83 theoretical modes.
These modes have nearby frequencies but different $k_z$ values, and in bin-by-bin cross-spectral estimation they contribute to the same complex cross-spectrum coefficient.
Therefore, the phase of this bin can be an effective phase produced by the complex-amplitude superposition of same-bin branches rather than the pure phase of the target branch.
Using the same probe geometries, $N_\mathrm{perseg}=2000$, and bin-by-bin MLE procedure, we compare two ablation conditions.
In the first condition, only the target branch is retained and the other 82 same-bin modes are removed.
In the second condition, all 83 theoretical modes in the bin are retained.
For both conditions, noiseless projected local likelihood profiles are constructed from analytic Fourier coefficients.

\begin{figure*}[t]
  \centering
  \includegraphics[width=\figwideone]{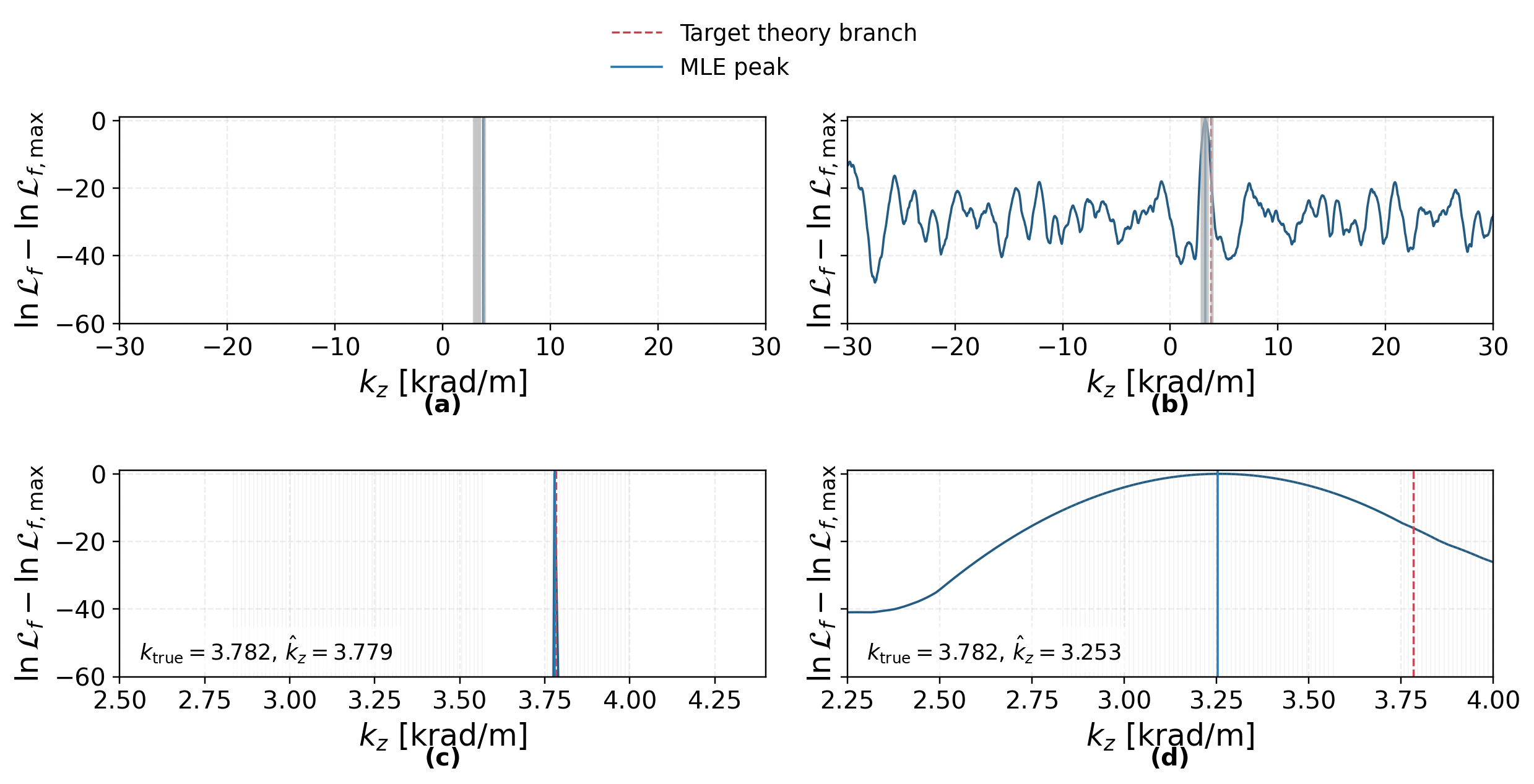}
  \caption{Focused ablation of multi-branch competition within a 50~kHz coarse frequency bin.
           Target branch: $f=1.727$~MHz, $k_z=3.782$~krad/m.
           Red dashed line: theoretical wavenumber; blue solid line: MLE main-peak; light-gray vertical lines: theoretical modes in the same FFT bin.
           (a,b) show the full wavenumber range; (c,d) zoom into the target branch.
           (a,c) With only the target branch retained, the MLE peak coincides with the theoretical value.
           (b,d) With all 83 same-bin theoretical branches retained, cross-spectral mixing reshapes the local peak structure and shifts the MLE main peak to lower $k_z$ --- an identity flip from the target branch to an effective mixed-branch peak.}
  \label{fig:bin_mixing_ablation}
\end{figure*}

When only the target branch is retained (Fig.~\ref{fig:bin_mixing_ablation}a,c), the MLE peak coincides with the theoretical value at $\hat{k}_z=3.779$~krad/m.
When all 83 same-bin modes contribute (panels b,d), the cross-spectral phase becomes a complex-vector superposition and the MLE main peak switches to $\hat{k}_z=3.253$~krad/m --- a clear identity flip from the target branch to an effective mixed-branch peak.
The ability of smaller $\delta f$ to improve multi-valued visibility therefore comes from both denser frequency-axis sampling and reduced same-bin branch mixing.

The top-$N$ output also includes alias candidates produced by phase wrapping, especially in the negative-$k_z$ region; isolating strict same-frequency multi-$k_z$ branches from these candidates requires additional frequency-continuity constraints, power weighting, growth-rate priors, or multimodal likelihood inversion.
This separates the inversion into two output levels: single-peak MLE characterizes the dominant dispersion ridge and its main folded structures, while top-$N$ local peaks retain the candidate information needed for subsequent multi-branch filtering.
We now return to resonance-order errors over acquisition duration and discuss how these likelihood mechanisms appear as convergence thresholds.

\subsection{Convergence thresholds of different resonance orders}

\begin{figure*}[t]
  \centering
  \begin{subfigure}[t]{\figtwoperrow}
    \centering
    \includegraphics[width=\linewidth]{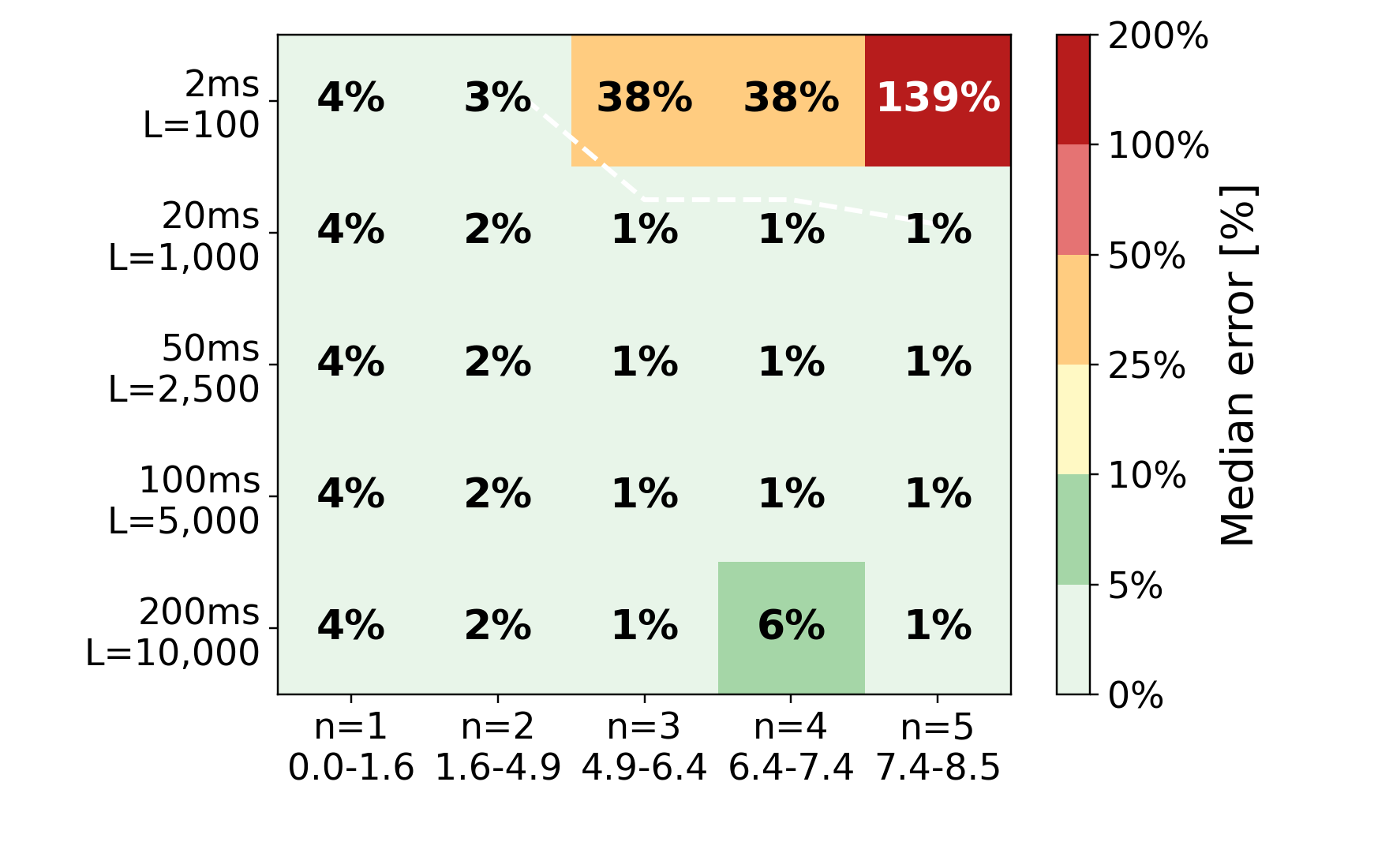}
    \caption{}
  \end{subfigure}\hfill
  \begin{subfigure}[t]{\figtwoperrow}
    \centering
    \includegraphics[width=\linewidth]{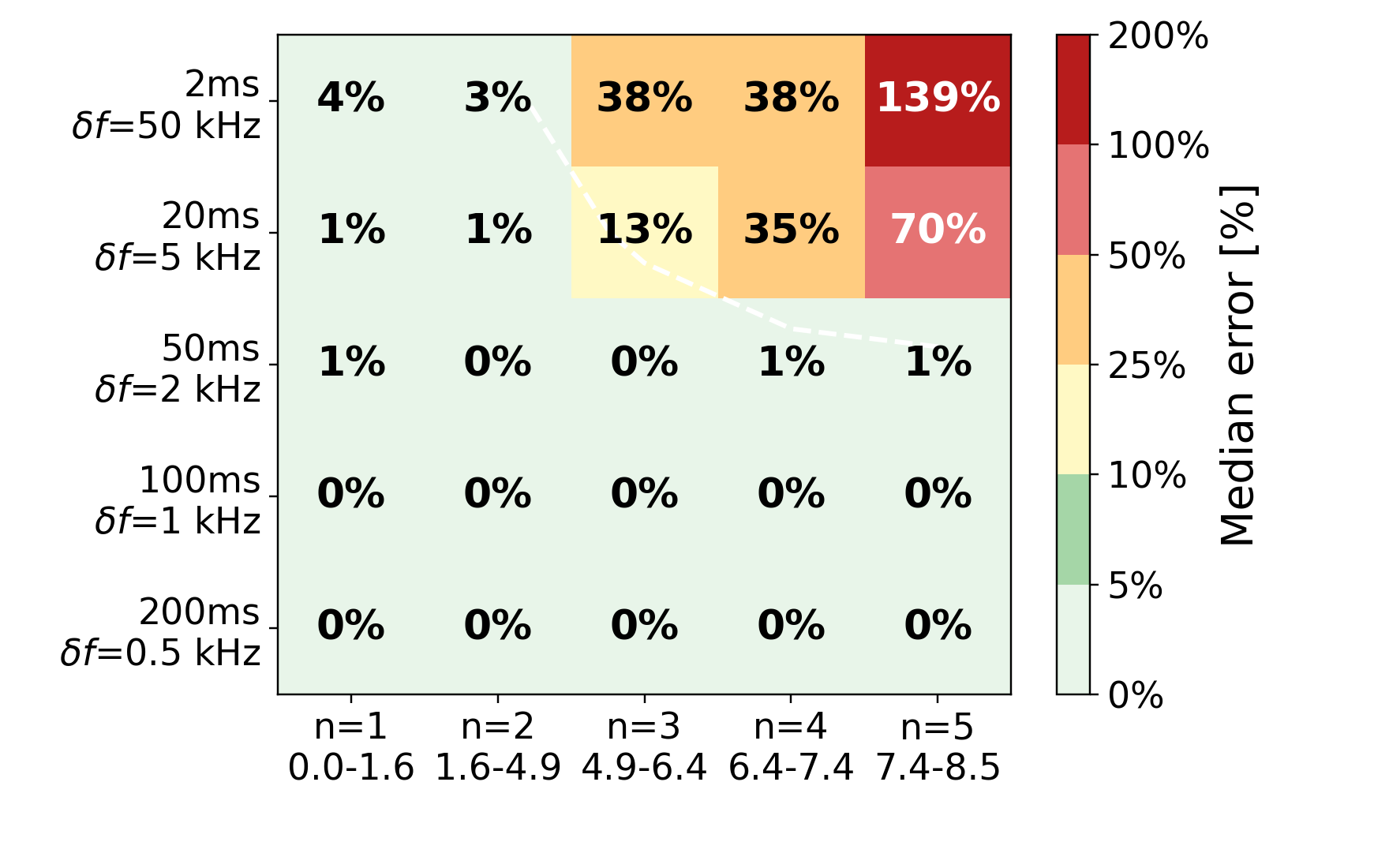}
    \caption{}
  \end{subfigure}
  \caption{Median relative error heat maps over resonance order and signal duration for the two schemes.
           Panel (a): Scheme 1 with fixed $\delta f=50$~kHz and increasing $L$.
           Panel (b): Scheme 2 with fixed $L=100$ and decreasing $\delta f$.
           Color: error magnitude; white dashed contour: 10\% convergence threshold.}
  \label{fig:heatmap_both}
\end{figure*}

Figure~\ref{fig:heatmap_both} compares the median relative error against resonance order and signal duration for the two schemes.

The comparison reveals a staged priority between $L$ and $\delta f$.

\textbf{(1) Short-duration stage: $L$ is the priority constraint.}
The fixed-$\delta f$ scheme shows that random outliers and spurious alias peaks decrease strongly as $L$ increases.
As summarized in Table~\ref{tab:threshold}, the first/second resonances converge within $L=100$ ($T=2$~ms), and the third to fifth resonances cross the threshold when $L\approx 1000$ ($T=20$~ms).
The error decreases through discrete threshold-like jumps rather than gradual improvement.
Together with Fig.~\ref{fig:likelihood_metrics}, this jump corresponds to the log-likelihood gap between the true branch and competing peaks crossing the selection boundary.
It is therefore a change in main-peak identity rather than a continuous narrowing of a single error band.
For high-order resonances, if $L$ is too small, the phase mean under finite realizations can still be dominated by noise and spurious peaks even when the theoretical dispersion location is clear.
Therefore, in short-duration or fast-insertion measurement windows, sufficient segment count $L$ should be secured before increasing $N_\mathrm{perseg}$ to pursue finer frequency resolution.

\begin{center}
  \refstepcounter{table}\label{tab:threshold}
  \scriptsize
  \textbf{TABLE~\thetable.} Minimum realization count required for convergence of each resonance order in Scheme 1 with fixed $\delta f$.
  \par\vspace{0.3em}
  \setlength{\tabcolsep}{1.2pt}
  \begin{tabular}{ccccc}
    \toprule
    Order & $f$ range & $k_z$ range & Error & Time \\
    $n$ & (MHz) & (krad/m) & at 2~ms &  \\
    \midrule
    1 & 0.0--1.6 & 0--4.6    & 4\%   & $<2$~ms  \\
    2 & 1.6--4.9 & 4.6--7.6  & 3\%   & $<2$~ms  \\
    3 & 4.9--6.4 & 7.6--13   & 38\%  & 2--20~ms  \\
    4 & 6.4--7.4 & 13--17    & 38\%  & 2--20~ms  \\
    5 & 7.4--8.5 & 17--22    & 139\% & 2--20~ms  \\
    \bottomrule
  \end{tabular}
\end{center}

\textbf{(2) Sufficient-duration stage: $\delta f$ is associated with local accuracy in folded regions.}
At fixed $L=100$, reducing $\delta f$ from 50~kHz to 2~kHz reduces the $n=4,5$ errors from 38\% and 139\% at 2~ms to 0.7\% and 0.6\% at 50~ms, meeting the 10\% criterion and becoming stable.
This convergence rate is comparable to Scheme 1, where $L$ increases at fixed $\delta f$.
However, Scheme 2 has smaller residual errors and narrower projected likelihood width near the folded extrema at 1.2 and 1.8~MHz.
Together with Figs.~\ref{fig:branch_gap_diagnostics}--\ref{fig:bin_mixing_ablation}, this shows that smaller $\delta f$ both increases the number of frequency samples and reduces mode competition within the same FFT bin --- the bin-mixing ablation confirms that multi-branch superposition is sufficient to shift the MLE main-peak.
Once $L$ is sufficient to form a stable joint-likelihood peak, the main benefit of further decreasing $\delta f$ is to compress local likelihood broadening in folded regions and reduce same-bin branch competition.

\textbf{(3) Combined design principle.}
$L$ and $\delta f$ should be regarded as two design parameters controlling different error sources.
Insufficient $L$ at short duration can cause global inversion failure.
Once $L$ exceeds the convergence threshold, $\delta f$ can become an important limitation on local errors near folded extrema.
Therefore, EDI spatial anti-aliasing experiments should follow the principle of first securing the segment count $L$ and then optimizing the frequency resolution $\delta f$.

\textbf{(4) Nonzero median errors of low-order resonances.}
The $n=1,2$ resonances retain median errors of 3--4\% at all durations.
This may reflect local errors from multi-$k_z$ branch mixing inside finite frequency bins near the first-resonance folded extrema.
Such local errors remain compatible with reconstruction of the overall EDI dispersion by multi-geometry MLE.

\section{Discussion}

The results identify a mechanism chain.
The nonmonotonic folded EDI dispersion changes the inverse frequency-to-wavenumber mapping from single-valued to multi-valued, so finite frequency bins allow several $k_z$ branches to contribute simultaneously to the cross-spectral phase.
The resulting joint likelihood contains the true peak, alias peaks produced by phase wrapping, and mixed peaks formed by folded branches.
The threshold-like convergence and local folded-point errors observed here are therefore consequences of this likelihood-topology change, with behavior distinct from ordinary gradual phase-variance reduction with sample number.

\subsection{Comparison with the IAW case}

The key difference between IAW and EDI is dispersion topology.
IAW in hollow-cathode plumes can usually be approximated as a monotonic linear dispersion, and Beall analysis, bispectral growth-rate measurements, and spatial anti-aliasing experiments have already been validated in this setting
\cite{liu2025,liu2025iepc,liu2024iepc,jorns2017}.
More generally, drift-driven electrostatic instabilities in nonmagnetized plasmas can transition among ion-acoustic, Langmuir, and Buneman-type branches as the electron--ion drift speed changes \cite{guo2013pst}.
In the IAW linear-dispersion case of Liu and Jorns \cite{liu2025}, the azimuthal-wavenumber estimation error $\sigma_{k_z}$ decreases gradually with data volume as

\begin{equation}
  \sigma_{k_z} \propto \frac{\sigma_\phi}{\sqrt{L} \cdot \Delta z},
  \label{eq:iaw_scaling}
\end{equation}

where $\sigma_\phi$ is the phase variance of a single measurement and $\Delta z$ is the baseline projection in the azimuthal direction.
The improvement is continuous and predictable.

The EDI results differ in two fundamental ways.

\textbf{(1) Threshold behavior replaces gradual improvement.}
The heat maps give a direct quantitative comparison.
For the $n=5$ resonance, the error drops from 139\% at 2~ms to 1.5\% at 20~ms and remains within about 1\% from 50 to 200~ms.
If a $1/\sqrt{L}$ law were extrapolated from 2~ms, the 20~ms case, with a tenfold larger $L$, would give $139\%/\sqrt{10}\approx44\%$.
The measured error has already decreased to 1.5\%, indicating a jump-like convergence between 2 and 20~ms rather than gradual improvement.

The physical origin of this threshold behavior is that the height difference between the true peak and the strongest spurious peak in the joint likelihood increases with $L$.
When this difference exceeds a critical value, MLE switches from the spurious peak to the true peak, producing a threshold-crossing transition instead of a continuous change.
In likelihood-topology terms, this jump corresponds to $\Delta\ln\mathcal{L}_f$ in Eq.~\eqref{eq:likelihood_gap} crossing from negative to positive, i.e., the true peak first exceeds the strongest competing peak.
High-order resonance convergence is therefore governed by a switch in the identity of the MLE main peak, rather than by a continuous decrease of phase variance alone.
A heuristic estimate follows from the number of aliased candidates and phase-mean statistics.
A true wavenumber $k_z$ above the Nyquist limit produces $N_\mathrm{wrap}\sim k_z/k_\mathrm{Nyq}$ phase-aliased candidates distributed across the wrapped phase interval, so the nearest spurious candidate is separated from the true peak by a phase increment $\Delta\phi\sim 2\pi/N_\mathrm{wrap}$.
The phase mean over $L$ realizations has random uncertainty $\sigma_\phi/\sqrt{L}$, where $\sigma_\phi$ is the single-realization phase standard deviation.
Resolving the true peak from its nearest spurious competitor in the joint likelihood requires $\sigma_\phi/\sqrt{L}\lesssim\Delta\phi$, which gives the critical realization count

\begin{equation}
  L^* \sim \left(\frac{\sigma_\phi}{2\pi}\right)^{2} N_\mathrm{wrap}^{2} \propto \left(\frac{k_z}{k_\mathrm{Nyq}}\right)^{2} .
  \label{eq:lstar_scaling}
\end{equation}

This estimate gives a scaling argument rather than an exact convergence duration because the separation between true and spurious peaks also depends on geometry combinations, noise level, multimodal mixing within frequency bins, and likelihood-peak width.
It explains why higher-order resonances usually require more data.

The implication for experiment design is that EDI diagnostics have a minimum realization threshold.
After this threshold is exceeded, further increasing $L$ gives diminishing returns for global convergence.
The heat maps show that 20~ms already brings all five resonances below the 10\% error threshold, while 50~ms provides a more conservative margin and the median error plateaus from 50 to 200~ms.
Experimental strategy should therefore first ensure that $L$ exceeds the convergence threshold of the target resonance order, then use additional total time to reduce $\delta f$ and improve local inversion quality near folded extrema.

In IAW scenarios, monotonically increasing $L$ is often sufficient for statistical convergence because the dispersion is single-valued.
In EDI, coarse $\delta f$ leaves local inversion errors near folded extrema even when $L$ is large, because multi-branch mixing within a frequency bin is independent of statistical noise.

\textbf{(2) Robustness of the joint likelihood to phase wrapping.}
At 200~ms, high-order resonances with $f>3$~MHz have large phase differences and degraded single-geometry phase statistics.
Nevertheless, the scatter plots still show good reconstruction.
This indicates that joint Bayesian estimation remains robust even when single-geometry phase statistics degrade, provided that the degradation differs among geometries and the joint constraints can still distinguish true and spurious peaks.
This feature is more prominent in the present EDI study than in the IAW case, where the wavenumber is smaller and phase wrapping is less severe.

\subsection{Phase inversion under folded EDI dispersion}

For a nearly monotonic linear dispersion, such as hollow-cathode IAW, the frequency-to-azimuthal-wavenumber mapping can be approximated as
\begin{equation}
  f\rightarrow k_z .
  \label{eq:iaw_mapping}
\end{equation}
In this case, the cross-spectral phase in each frequency bin is usually contributed by a single dominant branch, and increasing $L$ makes the phase mean closer to a conventional statistical-average problem.
For EDI, the dispersion folds near cyclotron resonances, and the inverse mapping becomes multi-valued:
\begin{equation}
  f\rightarrow\{k_{z,1},k_{z,2},\ldots\}.
  \label{eq:edi_mapping}
\end{equation}

The cross spectrum within a finite frequency bin no longer necessarily corresponds to a single wavenumber branch.
For probe baseline $i$ and frequency bin $j$, the mixed complex cross spectrum can be written schematically as
\begin{equation}
  C_i(f_j)\sim \sum_{m\in j} A_m^2
  \exp\!\left[-\mathrm{i}k_{z,m}\Delta z_i\right],
  \label{eq:cross_spectrum_mixture}
\end{equation}
Here $C_i(f_j)$ is the complex cross spectrum of probe baseline $i$ at frequency-bin center $f_j$, $m\in j$ denotes modes falling in that bin, $A_m$ and $k_{z,m}$ are the amplitude and azimuthal wavenumber of mode $m$, and $\Delta z_i$ is the effective azimuthal baseline projection along the propagation direction.
The observed phase is
\begin{equation}
  \Delta\Theta_i(f_j)=\arg C_i(f_j).
  \label{eq:observed_phase_mixture}
\end{equation}
This phase is generally a mixed phase determined jointly by the branches with the largest power and the most consistent cross-geometry phase alignment within the bin.
If one branch dominates a bin, single-peak MLE remains valid.
If several branches contribute comparably, likelihood peaks broaden, split, or switch identity.
If a spurious alias peak happens to align across multiple geometries, MLE can undergo a discrete misidentification jump.
EDI phase-inversion errors are therefore likelihood-topology problems involving the true peak, spurious peaks, and folded-branch mixed peaks, in addition to Gaussian phase-noise effects.

\subsection{Quantitative constraints for experimental design}

Based on the likelihood-topology analysis above, sampling-window design can be translated into three coupled but distinct constraints.

\textbf{Sampling-rate constraint.}
The sampling rate should satisfy $f_s>2f_\mathrm{max}\approx17$~MHz; the present 100~MHz value provides a conservative margin against temporal aliasing.

\textbf{Duration constraint.}
If all five resonances are targeted, $T_\mathrm{total}\gtrsim20$~ms ($L\gtrsim1000$) per angle-spacing configuration is sufficient to cross the 10\% error threshold; 50~ms ($L\gtrsim2500$) provides a more conservative margin.
The total experimental time across all 25 configurations must also include stage motion and configuration adjustment.

\textbf{Frequency-resolution constraint.}
If the goal is threshold-like recovery of the dominant dispersion ridge, $\delta f=50$~kHz is sufficient once $L$ is large enough.
If the goal includes local accuracy near folded extrema and visibility of multi-valued branches, $\delta f$ should be reduced to the kHz range.
The fixed-$L=100$ test shows that all five resonances meet the 10\% median-error criterion at $\delta f=2$~kHz, with further folded-region narrowing at 0.5~kHz.

These constraints control distinct error mechanisms: sampling rate controls temporal aliasing, $L$ controls random phase statistics and spurious-peak suppression, and $\delta f$ controls local frequency resolution near folded extrema.
Thus the practical priority is to secure the minimum $L$ required by the target resonance order before reducing $\delta f$ to resolve folded-region branch mixing.
Recent reviews of three-dimensional PIC studies emphasize that EDI dynamics and the associated fluctuation-driven transport are intrinsically three-dimensional \cite{zhao2026pst}.
Because the benchmark imposes azimuthally propagating modes with $k_y=0$ and stationary synthetic amplitudes, the numerical thresholds should be interpreted as design guidance for an azimuthally dominated EDI test rather than as universal values for arbitrary oblique or nonstationary spectra.

\section{Conclusions}

This work numerically examined the likelihood topology, error sources, and applicability limits of spatial anti-aliasing inversion for folded EDI dispersion in Hall thrusters.
The main conclusions are as follows.

\textbf{(1) Extension of measurable range.}
For 25 geometries, $f_s=100$~MHz, and SNR~=~10, the two-dimensional $(k_y,k_z)$ search recovers the dominant likelihood ridges associated with the first to fifth EDI resonances in 200~ms data.
The measurable wavenumber range is extended from $k_\mathrm{Nyq}=1$~krad/m to 22~krad/m, about 22 times the conventional Beall limit.
This shows that multi-geometry phase constraints can break the single-baseline Nyquist limit, although the recovered object is primarily the dominant likelihood ridge.

\textbf{(2) Likelihood-peak competition mechanism.}
The convergence of high-order EDI resonances is essentially a competition between the true peak and spurious alias peaks.
As $L$ increases, the log-likelihood gap of the true peak over competing peaks increases, and MLE can switch from a spurious peak to the true peak.
The error decrease is therefore threshold-like rather than the $1/\sqrt{L}$ gradual convergence common in linear IAW dispersion.
For the present parameters, $n=1,2$ satisfy the threshold within 2~ms and $n=3$--5 cross it at about 2--20~ms.

\textbf{(3) Frequency-resolution mechanism in folded regions.}
Near folded extrema, finite $\delta f$ makes several $k_z$ branches contribute to the cross-spectral phase, causing peak broadening, peak switching, and local errors.
Reducing $\delta f$ compresses the likelihood width in folded regions, reduces branch competition within the same frequency bin, and increases the chance that multi-valued branches become main peaks or low-rank candidate peaks at adjacent fine frequency points.
Thus $\delta f$ controls local broadening and branch mixing in folded regions, complementing the role of $L$ in spurious-peak suppression.

\textbf{(4) Algorithmic applicability limit.}
The two-dimensional single-peak MLE recovers the dominant dispersion ridge and its main folded structures for the present azimuthally dominated benchmark with prescribed $k_y=0$ ground truth.
Strict enumeration of same-frequency multi-$k_z$ branches requires additional multimodal processing, since top-$N$ local peaks include both true folded-branch and alias candidates.
Oblique propagation, spatial nonuniformity, nonlinear saturation, and multimodal coexistence in real experiments require further experimental validation.

Therefore, for Hall-thruster EDI wave-probe diagnostics, sampling design should first meet the minimum $L$ required by the target resonance order and then reduce $\delta f$ within the allowed dwell time, so that both spurious-peak suppression and local folded-region accuracy are maintained.

\section*{Acknowledgment}

The authors acknowledge the support from National Natural Science Foundation of China (Grant Nos.~52472403 and U25B20235).

\section*{Data Availability Statement}

The data that support the findings of this study are available from the corresponding author upon reasonable request.

\end{document}